\documentstyle[12pt]{article}
\begin{document}
\pagestyle{empty}
\hfill{ANL-HEP-PR-95-82}
\vskip 1cm
\centerline{\Large \bf The Perturbative Resummed Series}
\centerline{\Large \bf for Top Quark Production in Hadron Reactions }
\vskip .5cm
\begin{center}
Edmond L. Berger and Harry Contopanagos\\
High Energy Physics Division\\
Argonne National Laboratory\\
Argonne, IL 60439\\
\end{center}
\vskip .5cm
\centerline{March 15, 1996}
\vskip .5cm
\centerline{\bf Abstract}
Our calculation of the total cross section for inclusive production of
$t\bar{t}$ pairs in hadron collisions is presented.  The principal ingredient
of the calculation is resummation of the universal leading-logarithm effects
of gluon radiation to all orders in the quantum chromodynamics coupling
strength, restricted to the region of phase space that is demonstrably 
perturbative.  We derive the perturbative regime of
the resummed series, starting from the principal-value resummation approach,
and we isolate the perturbative domain in both moment space and, upon
inversion of the corresponding Mellin transform, in momentum space.  We show
that our perturbative result does not depend on the manner non-perturbative or 
infrared effects are handled in principal-value resummation.  We treat both 
the quark-antiquark and gluon-gluon production channels consistently in the 
$\overline{{\rm MS}}$ factorization scheme.  We compare our method and results 
with other resummation methods that rely on the choice of infrared cutoffs.  
We derive the renormalization/factorization scale dependence of our resummed
cross section, and we discuss factorization scheme dependence and remaining
theoretical uncertainties, including estimates of possible non-perturbative
contributions.  We include the full content of the exact next-to-leading order
calculation in obtaining our final results. We present predictions of the
physical cross section as a function of top quark mass in proton-antiproton
reactions at center-of-mass energies of 1.8 and 2.0 TeV.  We also provide the
differential cross section as a function of the parton-parton subenergy.

\newpage
\section{Introduction}
\pagestyle{plain}

In hadron interactions at collider energies, the main production mechanisms
for inclusive top-antitop quark ($t\bar{t}$) production, as modeled in
perturbative quantum chromodynamics (pQCD), involve parton-parton collisions.
The perturbative series begins at second order, ${\cal O}(\alpha_s^2)$, in the
strong coupling strength, $\alpha_s$.  At this order, the parton subprocesses
are
\begin{equation}
i+j\rightarrow t+\bar{t},
\label{zero}
\end{equation}
where the initial partons $i,j$ are either a quark-antiquark ($q\bar{q}$) or a
gluon pair ($g g$). In higher orders, gluons are radiated in these two
production channels, and there are additional production channels, such as
$q g \rightarrow t \bar{t} q$.  The gluonic radiative corrections
to the lowest-order channels create large enhancements of the partonic cross
sections near the top pair production threshold\cite{ref:dawson,ref:vneerven}.
The magnitude of the ${\cal O}(\alpha_s^3)$ corrections implies that fixed-order
perturbation theory will not necessarily provide reliable quantitative
predictions for ($t\bar{t}$) pair production at Fermilab Tevatron energies.  A
resummation of the effects of gluon radiation to all orders in perturbation
theory is called for in order to improve the reliability of the theory.  This
was the main motivation for the published resummation calculations of
$t\bar{t}$ production\cite{ref:laeneno,ref:laenent,ref:edpapero}.

In a prior
paper, we presented a brief exposition of our method of resummation and its
application to $t\bar{t}$ production\cite{ref:edpapero}. Our guiding principle
is the all-orders resummation of the universal leading-logarithm effects of
initial-state radiation, restricted to the region of phase space which is
manifestly {\it perturbative}.  Our purpose in this paper is to expand upon
our earlier theoretical and phenomenological discussion.  We provide a
detailed exposition of our approach, highlight the similarities and
differences with Ref.~\cite{ref:laeneno,ref:laenent}, and show that our
predictions are independent of the particular regularization of the infrared and
non-perturbative regions imposed by principal-value resummation (PVR). We
present predictions of the inclusive top quark production cross section as a
function of top quark mass in proton-antiproton reactions at center-of-mass
energies $\sqrt{S}= $ 1.8 and 2.0 TeV, and in proton-proton reactions at the
energies of the CERN Large Hadron Collider (LHC).  We demonstrate that the
renormalization/factorization scale dependence of our resummed cross sections is
very modest.  At the end, we speculate on modeling the unknown dynamics of the
non-perturbative region and its possible contribution to the cross section.

In ``infrared safe" processes, such as the total cross section for
$e^+e^-$-annihilation to hadrons, there is only one physical scale that
characterizes the perturbative cross section, namely the total energy
of the electroweak initial state. In hadron-hadron scattering, according to
the factorization theorem of pQCD, the cross sections for the partonic
sub-processes are the main object of theoretical calculations.
In most hard-scattering processes, however, partonic cross sections are not
free of singularities.  Initial-state hadronic interactions require that a
non-trivial ``mass-factorization" be performed in order to absorb a singular
part of these cross sections into a redefinition of the long-distance parton
distribution functions in a universal, process-independent way.  Once
mass-factorization is performed, the object of theoretical interest is the
short-distance part of the partonic cross sections (the ``hard part").  The
hard part usually depends on more than one momentum scale, typically
because of gluon radiation.  The hard part is calculated as a series in the
strong coupling strength of pQCD, and the domain of applicability of the series
is a function of {\it all} pertinent momentum scales. In determining the
region of applicability of this series, one encounters difficulties including,
but not limited to, infrared renormalons\cite{ref:renormalons}.  Most
discussions (not necessarily in agreement) on the range of validity of pQCD
have focused on momentum-scaling properties of Green's
functions\cite{ref:momentums,ref:beneke}, but color factor constants are
an equally important aspect of the series and should be taken seriously
taken into account.  Both can contribute to the breakdown of pQCD.  Although
the former have a more ``dynamical" appearance because they depend on momentum,
both create difficulties that are combinatoric in nature. Infrared renormalons
create factorial growth associated with massless vacuum polarization diagrams,
and large constants enhance the strength of the partonic interactions
according to their corresponding fields in color space. Among the many virtues
of Ref.~\cite{ref:laeneno} is the demonstration that large multiplicative color
factors are indeed extremely important in realistic situations. Their effects
transcend standard operator-product-expansion (OPE) arguments to the effect
that non-perturbative effects are suppressed by at least $\Lambda^2/m^2$ or
approaches that are usually formal and model-dependent\cite{ref:beneke}.  We
treat this problem in detail in the present paper from the realistic
viewpoint of applying resummation to a specific reaction of significant
phenomenological importance.  From an operator point of view, however, this is
an issue that deserves further investigation, to be addressed in a future
publication\cite{ref:future}.

Top quark production, Eq.~(\ref{zero}), embodies all the issues mentioned
above. An instructive illustration can be found in the ${\cal O}(\alpha_s^3)$
calculation of the partonic hard part, or loosely speaking, the ``partonic
cross section" for heavy flavor production\cite{ref:dawson}.  Figures 12 and
13 of the first paper in Ref.~\cite{ref:dawson} show that the
${\cal O}(\alpha_s^3)$ hard part is much larger than the leading-order
${\cal O}(\alpha_s^2)$ hard part {\it in specific regions of partonic
subenergy}.  Analogous results are shown here in Fig.~7, to be discussed in
section 5.  Not all of the range of the partonic subenergy can be considered
part of the perturbative domain {\it despite the fact} that, at
finite orders, the resulting radiative corrections are integrable throughout
that phase space and yield finite inclusive predictions. Dominance of the
higher-order contributions in the $q\bar{q}$ and $g g$ channels near threshold,
and in the $g g$ channel at large values of the subenergy, feeds back (albeit
not as enhanced) to the physical cross sections obtained after convolution
with parton densities.  In this paper, we limit our attention to processes in
which the near-threshold region is the most important influence.  At Fermilab
Tevatron collider energies, the next-to-leading order enhancement of the top
quark cross section is approximately $25\%$ relative to the leading-order
value.  For comparison, at a scale specified by the mass $m$ of the top quark,
$\alpha_s(m)\simeq 10\%$.  Top quark production at the Tevatron is one that
involves {\it multiple} QCD scales.

Dominance of the next-to-leading order contributions to top quark production
near threshold in the $q\bar{q}$ and $g g$ channels is mentioned earlier in
this Introduction as a primary motivation for theoretical study of resummation
of the effects of gluon radiation to all orders.  One method of resummation was
implemented for $t\bar{t}$ pair production some time
ago\cite{ref:laeneno,ref:laenent}.  In most resummation methods, the
threshold corrections are exponentiated into a function of the QCD running
coupling strength, $\alpha_s$, evaluated at a variable momentum scale which is
a measure of the radiated momentum. To recover the inclusive cross section,
one then integrates over the available radiation phase space down to the
partonic threshold.  An inherent ambiguity and limitation of the method of
Ref.~\cite{ref:laeneno} is the introduction of undetermined infrared (IR)
cutoffs.  The need for these cutoffs arises because the kinematically allowed
region meets the Landau pole of the QCD running coupling. More generally, the
exponentiation of logarithmic corrections results in essential singularities,
whether expressible as singularities of the running coupling or not,
while a finite-order expansion in terms of $\alpha_s$ at a fixed hard scale
exhibits a polynomial (and hence integrable) dependence on the logarithms.
Since the IR cutoff dependence is exponentiated in Ref.~\cite{ref:laeneno}, the
sensitivity of their predicted cross sections to the value of the cutoffs
is very significant numerically, especially when multiplicative color factors
are large, as in the $gg$ production channel.

An advantage of our resummation approach\cite{ref:edpapero} is that it does not
require arbitrary IR cutoffs. We use the principal-value resummation (PVR)
technique\cite{ref:stermano} to by-pass the Landau poles and associated
renormalon singularities, and we obtain a mathematically unambiguous expression
for the resummation exponent in moment space. The PVR procedure allows us to
identify the {\it perturbative} properties of the resummation exponent and to
separate these from the {\it non-perturbative} behavior of the exponent. The
perturbative properties are obtained through a short-distance asymptotic
approximation valid in a specific region of moment space.  The separation is
in turn used to derive a perturbative regime for the hard part itself in moment
space, which is more restricted due to large color factors and exponentiation.
The inversion of the Mellin transform\cite{ref:stermanz} provides
the hard part directly in momentum space, and we obtain a resummed perturbative
expression for the partonic cross section, along with the corresponding
determination of the perturbative regime in momentum space.  This new method
should be contrasted with the resummation approach of Ref.~\cite{ref:laeneno}
where explicit use of IR cutoffs makes a perturbative separation impossible.

It may be argued that adoption of the PVR technique is the adoption of a model
and that the IR cutoff dependence of Ref.~\cite{ref:laeneno} is merely
concealed.  We argue that this is not the case precisely because we apply our
method only in a well defined perturbative region.  To the extent that PVR is
a model, it is a model only for the non-perturbative region that we do not
include.  The main physical issue on which we focus is not the finiteness
PVR regularization imposes.  Rather, it is how this finiteness {\it helps us
to probe} the asymptotic properties of the perturbative cross section. Therein
lies part of the usefulness of PVR, even though it is possible
that the regularization it imposes throughout phase space, including the
non-perturbative regime, may be physically significant.
For example, PVR applied to the Drell-Yan process at fixed target energies
is in excellent agreement with experiment\cite{ref:alvero}.
We will discard the cross
section in the non-perturbative region as model-dependent, notwithstanding its
finiteness in PVR.

The perturbative regime should be independent of infrared regulators,
including the principal-value regulator.  To demonstrate this independence, we
show in detail in this paper how one may obtain the same
expressions as ours for the perturbative resummed cross section, independently
of any regularization.  To do so, we must make two physically plausible
assertions about the perturbative behavior of the resummation exponent in
moment space, assertions that are natural ingredients of the PVR exponent.

Our resummation includes the {\it leading} large threshold logarithmic
contributions to all orders in perturbation theory. These contributions are
interpreted to be universal, characteristic of the initial states that produce
the hard scattering.  They are common with other hard scattering processes,
notably massive lepton-pair ($l\bar{l}$) production, the Drell-Yan process.
A complete ${\cal O}(\alpha_s^2)$ calculation is available for the Drell-Yan
process, meaning that the logarithmic structure is known explicitly in that
case to one order in perturbation theory higher than for $t\bar{t}$
production.  Final-state effects, which differentiate among hard-scattering
processes, produce subleading logarithmic structures that are not universal
and should not be included in a universal resummation approach. It is possible,
and even probable, that resummation of subleading large logarithms from
final-state and interference effects can be accomplished, but such a
resummation will be process-dependent. For the application we have in mind,
its numerical effects will be subleading in nature.  We consider the
resummation of final-state logarithms beyond the scope of this work.

An outline of this paper is follows.  In section 2 we present the kinematics of
the inclusive cross section for top quark production and the form of its
resummed version. We include a general discussion on counting powers of
logarithms for the hard part at finite orders and for its resummed version.
In section 3 we present the perturbative properties of the PVR resummation
exponent, $E$, in moment space, in both the $\overline{{\rm MS}}$ and DIS
factorization schemes, as well as its renormalization-group (RG) invariance
properties.  We consider a formally identical exponent but without
regularization, and we examine its factorial growth in moment space in order
to compare the resulting perturbative asymptotic approximation
{\it at fixed moments} with the one resulting from PVR.  The two perturbative
approximations are in agreement with each other, but the one without the PVR
prescription does not directly constrain the moment variable since it derives
from an unregularized expression that exists only formally.  We offer
additional physically motivated criteria one could use in this latter approach
to constrain the moment variable, and we show that the resulting constraints
are similar to those imposed by PVR.  Using a heuristic argument, we describe
how one may estimate the perturbative regime for the exponentiated hard part
itself in moment space.

In section 4 we discuss the universality of the threshold logarithmic
corrections in comparison with massive lepton-pair ($l\bar{l}$) production,
using threshold asymptotics and the results of the complete
${\cal O}(\alpha_s^2)$ calculation of heavy quark
production\cite{ref:dawson,ref:vneerven}. We examine the
detailed structure of the hard-scattering function in the gluon radiation phase
space (momentum space) and the corresponding determination of the perturbative
regime in that space.  The hard-scattering function is proportional to
${\rm exp}(E)$, which resums the threshold corrections directly.  The inversion
of the Mellin transform produces a hart part in momentum space that includes a
series of subleading structures deriving from $E$\cite{ref:stermanz}.  It is
this latter expression that is best suited for the determination of the
perturbative regime in momentum space. The two determinations of the
perturbative regime for the cross section in moment and momentum space are in
good agreement with each other. We conclude that the resummed perturbative
partonic cross section derived from the PVR approach is independent of the
PVR regularization and is valid in a well-established perturbative regime.
It is derived basically from simultaneous minimization of the factorial growth
of IR renormalons {\it and} color-factor combinatorics in the exponentiated
hard part.

In section 5 we discuss the numerical properties and phenomenological behavior
of the resummed partonic and physical cross sections, for both the $q\bar{q}$
and $g g$ production channels, in the $\overline{{\rm MS}}$ and DIS
factorization schemes.  Predictions are presented in the form of both figures
and tables of the inclusive cross section for top quark production as a
function of top mass in proton-antiproton collisions at
$\sqrt{S}= $ 1.8 and 2.0 TeV.  We display the factorization/renormalization
scale dependence of the physical cross sections and provide extensive
comparisons with the corresponding quantities of Ref.~\cite{ref:laeneno}.  We
address the issue of perturbative theoretical uncertainties in both our
approach and that of Ref.~\cite{ref:laeneno,ref:laenent}.  In section 6 we
present estimates of non-perturbative uncertainties based on physically
motivated assumptions about the behavior of the cross section in that regime.
We summarize our conclusions in section 7.

\section{Production kinematics and resummation}

In this section we begin with expressions for the partonic and physical
cross sections in finite orders of QCD perturbation theory and the associated
kinematics of $t\bar{t}$ production. Subsequently, we present a resummed
expression for the partonic cross section in PVR, based on universality of the
leading threshold corrections with those in massive lepton-pair ($l\bar{l}$)
production.  In this paper, we use upper-case $S$ to denote the square of the
total energy in the hadron-hadron system and lower-case $s$ for the square of
the energy in the partonic system.

\subsection{Next-to-leading-order cross section and kinematics}

We start with the next-to-leading-order one-particle inclusive partonic
differential cross section in the DIS factorization scheme.  We use the
notation $\alpha(\mu)\equiv \alpha_s(\mu)/\pi$, where $\mu$ is the common
renormalization/factorization hard scale of the problem. A perturbative
quantity $R$ is expanded as $R^{[i]}=\sum_{j=0}^i\alpha^j R^{(j)}$, where
$R^{(j)}$ is the order $\alpha^j$  radiative correction to $R$,
{\it above the leading order}.  Unless otherwise specified,
$\alpha\equiv \alpha(\mu=m)$ where $m$ is the mass of the top quark.
Following the notation of Ref.~\cite{ref:laeneno} for the subprocess
\begin{equation}
i(k_1)+j(k_2)\rightarrow t(p_1)+{\bar t}(p_2)+g(k),
\label{subprocess}
\end{equation}
and defining the partonic invariants
\begin{equation}
s=(k_1+k_2)^2,\ t_1=(k_2-p_2)^2-m^2,\ u_1=(k_1-p_2)^2-m^2,\ s_4=s+t_1+u_1,
\label{invariants}
\end{equation}
we express the partonic differential cross section as
\begin{equation}
m^2s^2{d^2\sigma_{ij}^{(1)}\over dt_1du_1}(s,t_1,u_1)=
2C_{ij}\bar{D}_1(s_4;\Delta)\sigma_{ij}^B(s,t_1,u_1) + R\ .
\label{two}
\end{equation}
In Eq.~(\ref{two}), $\sigma_{ij}^B(s,t_1,u_1)$ is the lowest-order Born cross
section, expressed in terms of three-particle final-state variables, and the
remainder $R$ stands for terms that do not contain leading logarithmic
corrections near threshold.  The quantity $C_{ij}$ is the color factor for the
$ij$ production channel, and
\begin{equation}
\bar{D}_1(s_4,m^2;\Delta)\equiv {m^2\over s_4}\ln\left({s_4\over m^2}\right)
\Theta(s_4-\Delta)+m^2\delta(s_4){1\over 2}\ln^2\left({\Delta\over m^2}\right)\ ,
\label{three}
\end{equation}
where $\Delta$ is understood in the distribution sense:
\begin{equation}
\lim_{\Delta\rightarrow 0}\int_0^y d(s_4/m^2)\phi(s_4/m^2)
\bar{D}_1(s_4,m^2;\Delta)\ .
\label{four}
\end{equation}
In Eq.~(\ref{invariants}), $s_4$ is a measure of the inelasticity
of the radiative process, and it is proportional to the gluon momentum $k$.
One can show that $s_4=2k\cdot p_1$ and hence, for soft gluons,
$s_4\rightarrow 0$.
The distribution $\bar{D}_1$ can be shown to be a ``plus" distribution.
Indeed, define $s_4/m^2\equiv 1-z$ and
\begin{equation}
D_1(z)\equiv\left({\ln(1-z)\over 1-z}\right)_+\ .
\label{irrelevanto}
\end{equation}
Then, for any smooth function $\phi$,
\begin{equation}
\int_{1-y}^1dz\phi(1-z)D_1(z)\equiv \int_{1-y}^1dz[\phi(1-z)-\phi(0)]
{\ln(1-z)\over 1-z}-
\phi(0)\int_0^{1-y}dz{\ln(1-z)\over 1-z}\ .
\label{irrelevantt}
\end{equation}
The right-hand-side can be written as
\begin{eqnarray}
& &\int_{1-y}^{1-\epsilon}dz\phi(1-z){\ln(1-z)\over 1-z}
-\phi(0)\int_{1-y}^{1-\epsilon}dz{\ln(1-z)\over 1-z}\nonumber \\
& &+\sum_{k=1}^\infty{\phi^{(k)}(0)\over k!}\int_{1-\epsilon}^1
dz(1-z)^{k-1}\ln(1-z)-\phi(0)\int_0^{1-y}dz{\ln(1-z)\over 1-z}\ .
\label{five}
\end{eqnarray}
The series tends to zero as $\epsilon\rightarrow 0$, and the result is
\begin{eqnarray}
\int_{1-y}^1dz\phi(1-z)D_1(z)& &=
\int_{1-y}^{1-\epsilon}dz\phi(1-z){\ln(1-z)\over 1-z}+\phi(0){1\over 2}\ln^2
\epsilon\nonumber \\
& &=\int_\epsilon^y {ds_4\over s_4}\phi(s_4/m^2)\ln\left({s_4\over m^2}\right)+
\phi(0){1\over 2}\ln^2\epsilon\ .
\label{six}
\end{eqnarray}
This is the result obtained also from Eqs.~(\ref{three}) and (\ref{four})
with the identification $\epsilon\equiv \Delta/m^2$.
Hence we have proved the identity $\bar{D}_1(s_4,m^2;\Delta)=D_1(z)$,
with $s_4/m^2\equiv 1-z$. The identification $s_4/m^2\equiv 1-z$ is suggestive
of the similarity of the present reaction, $t\bar{t}$ production, with the
Drell-Yan process\cite{ref:stermano,ref:alvero} where $z=Q^2/ s$ is the
fraction of the squared invariant energy carried by the dilepton pair.
The threshold is at $z\rightarrow 1$, as it is in our case as well.  We use $z$
rather than $s_4$ in this paper to stress the similarity between these two
reactions.

We can write the differential cross section, including the Born term, as
\begin{equation}
m^2s^2{d^2\sigma_{ij}^{[1]}\over dt_1du_1}(s,t_1,u_1)=
\Biggl\{\delta(1-z)+\alpha 2C_{ij}\left({\ln(1-z)\over 1-z}\right)_+\Biggr\}
\sigma_{ij}^B(s,t_1,u_1)\ .
\label{seven}
\end{equation}
If we compare the above expression and the corresponding one for the Drell-Yan
process, we may verify that the logarithmic structure is identical through
next-to-leading order.  It is important to note here that the exact
calculation of the cross section for $t\bar{t}$ production, including the
remainder term $R$ in Eq.~(\ref{two}), contains subleading terms, such as
$(1/(1-z))_+$, as well as constants. These structures are not
common to the Drell-Yan reaction, and they cannot be resummed as part
of initial-state radiation only. As in Ref.~\cite{ref:laeneno}, we
disregard these subleading structures in our resummation.  We demonstrate the
non-universal character of these subleading corrections in section 4.

Following Ref.~\cite{ref:laeneno} we integrate Eq.~(\ref{seven}) over the
whole partonic phase space. Using the appropriate kinematic bounds, we obtain
\begin{equation}
\sigma_{ij}^{[1]}(\eta,m^2)=\int_{1-4(1+\eta)+4\sqrt{1+\eta}}^1
dz\Biggl\{\delta(1-z)+\alpha 2C_{ij}\left({\ln(1-z)\over 1-z}\right)_+\Biggr\}
\bar{\sigma}_{ij}^B(\eta,z,m^2)\ ,
\label{eight}
\end{equation}
where
\begin{equation}
\bar{\sigma}_{ij}^B(\eta,z,m^2)\equiv {\sqrt{(s/m^2-1+z)^2-4s/m^2}\over 2s^2}
\int_{-1}^1d\cos\theta\sigma_{ij}^B(s,t_1,u_1) \ .
\label{nine}
\end{equation}
The kinematic transformations
\begin{eqnarray}
& &t_1=-{1\over 2}m^2
\left(s/m^2-1+z-\sqrt{(s/m^2-1+z)^2-4s/m^2}\cos\theta\right)\nonumber \\
& &u_1=-{1\over 2}m^2
\left(s/m^2-1+z+\sqrt{(s/m^2-1+z)^2-4s/m^2}\cos\theta\right)\nonumber \\
& &s=4m^2(1+\eta)\ .
\label{kinematictrans}
\end{eqnarray}
are used to obtain Eqs.~(\ref{eight}) and (\ref{nine}).  The lower limit of
integration in Eq.~(\ref{eight}) is derived from  the kinematics of
Eqs.~(\ref{subprocess}) and (\ref{invariants}), namely $s_4\le s-2m\sqrt{s}$.

As in the Drell-Yan case, one may eliminate the delta function and
plus-distribution in Eq.~(\ref{eight}) by integrating by parts. Using
the identity
\begin{equation}
\int_z^1dx(f(x))_+=-\int_0^zdxf(x)\ ,
\label{irrelevantth}
\end{equation}
and the kinematic constraint
\begin{equation}
\bar{\sigma}_{ij}^B\biggl(\eta,z=1-4(1+\eta)+4\sqrt{1+\eta}),m^2\biggr)=0\ ,
\label{irrelevantf}
\end{equation}
we arrive at the following expressions for the cross sections.  In the DIS
factorization scheme,
\begin{equation}
\sigma_{ij}^{[1]}(\eta,m^2)=\int_{1-4(1+\eta)+4\sqrt{1+\eta}}^1dz
\left\{1+\alpha C_{ij}\ln^2(1-z)\right\}\sigma'_{ij}(\eta,z,m^2)\ ,
\label{twelve}
\end{equation}
and, in the $\overline{{\rm MS}}$ scheme,
\begin{equation}
\sigma_{ij}^{[1]}(\eta,m^2)=\int_{1-4(1+\eta)+4\sqrt{1+\eta}}^1dz
\left\{1+\alpha 2 C_{ij}\ln^2(1-z)\right\}\sigma'_{ij}(\eta,z,m^2)\ .
\label{twelvep}
\end{equation}
\begin{equation}
\sigma'_{ij}(\eta,z,m^2)\equiv {d\over dz}\bar{\sigma}_{ij}^B(\eta,z,m^2)\ .
\label{fourteen}
\end{equation}
The explicit expression for the derivative of the Born cross section
in the $q\bar{q}$ channel is
\begin{equation}
\sigma'_{q\bar{q}}(\eta,z,m^2)={2\over 3}C_F\pi^3\alpha^2{\tau\over s} x(z)
\left\{\sqrt{x^2(z)-4\tau}+
{2\tau\over \sqrt{x^2(z)-4\tau}}\right\}\ ,
\label{fifteenp}
\end{equation}
where
\begin{equation}
x(z)\equiv 1-(1-z)\tau\ ,
\label{irrelevantfi}
\end{equation}
$\tau\equiv m^2/s=[4(1+\eta)]^{-1}$, and $C_F=4/3$.  In the $gg$ channel,
\begin{eqnarray}
& &\sigma'_{gg}(\eta,z,m^2)={3\over 16}\pi^3\alpha^2C_F{\tau\over s}
\Biggl[C_F\Biggl\{-\biggl(1-{4\tau\over x^2(z)}\biggr)^{3/2}\nonumber \\
& &+\biggl(1-{4\tau\over x^2(z)}+{24\tau^2\over x^4(z)}\biggr)\ln y(z)
+\biggl(2+{4\tau\over x^2(z)}-{32\tau^2\over x^4(z)}\biggr){1\over \sqrt{1-
{4\tau\over x^2(z)}}}\Biggr\}\label{fifteenpp} \\
& &+C_A\Biggl\{-\biggl(x^2(z)+{52\over 3}\biggr)\sqrt{1-{4\tau\over x^2(z)}}
-{4\tau^2\over x^2(z)}\ln y(z)-\biggl({4\tau\over 3}-{4\tau^2\over 3x^2(z)}
\biggr){1\over \sqrt{1-{4\tau\over x^2(z)}}}\Biggr\}\Biggr],\nonumber
\end{eqnarray}
where $C_A=3$ and
\begin{equation}
y(z)\equiv {1+\sqrt{1-{4\tau\over x^2(z)}}\over 1-\sqrt{1-{4\tau\over x^2(z)}}}
\ .\label{fifteenppp}
\end{equation}

\subsection{Resummation and power counting}

The large logarithms near threshold, both at finite-orders
and in resummation, play a major role in our considerations.  In this section,
we provide a general description that establishes the notion of leading and
subleading logarithmic structures and the way we use this terminology.
Generalizing the notation of Eqs.~(\ref{twelve}) and (\ref{twelvep}),
we write the partonic cross section resulting from a finite-order perturbative
calculation as
\begin{equation}
\sigma^{[k]}(\eta,m^2)=\int_{1-4(1+\eta)+4\sqrt{1+\eta}}^1dz
H^{[k]}(z,\alpha)\sigma'_{ij}(\eta,z,m^2)\ .
\label{ntwelve}
\end{equation}
Here
\begin{equation}
H^{[k]}(z,\alpha)=\sum_{m=0}^k\alpha^m\sum_{l=0}^{2m}c(l,m)x_z^l\ ,\
{\rm with}\ x_z\equiv \ln\biggl({1\over 1-z}\biggr) \ ,
\label{explano}
\end{equation}
and $c(l,m)$ are calculable numerical coefficients. In this representation,
the classification of the logarithmic structures is obvious. Leading
logarithms are the monomials proportional to $c(2m,m)$, first subleading
logarithms the monomials proportional to $c(2m-1,m)$, etc. Since we are
concerned with threshold enhancements, $z\rightarrow 1$ or, equivalently,
$x_z\rightarrow +\infty$, the nomenclature is related directly to the
numerical importance of the corresponding logarithmic structures in the cross
section.

At finite orders, the logarithmic structures are integrable near threshold,
and their contributions to the cross section are finite, provided the Born
cross sections are also integrable (as is indeed the case).
For purposes of demonstration, let us assume that the Born cross section
behaves as $(1-z)^{\nu}$, $-1<\nu$. Using
\begin{equation}
x_z^l=\ln^l\biggl({1\over 1-z}\biggr)=(-1)^l
\lim_{\epsilon\rightarrow 0}\left({\partial
\over \partial \epsilon}\right)^l(1-z)^\epsilon \ ,
\label{irrelevants}
\end{equation}
and denoting $L(\eta)\equiv 1-4(1+\eta)+4\sqrt{1+\eta}$, we find
\begin{eqnarray}
H^{(l,m)}(\eta,\alpha)&\equiv&\alpha^mc(l,m)\int_{L(\eta)}^1dz(1-z)^{\nu}x_z^l=
\alpha^mc(l,m)(-1)^l\lim_{\epsilon\rightarrow 0}
\left({\partial
\over \partial \epsilon}\right)^l
\int_{L(\eta)}^1dz(1-z)^{\nu+\epsilon}\nonumber \\
& &=\alpha^mc(l,m)(-1)^l\lim_{\epsilon\rightarrow 0}\left({\partial
\over \partial \epsilon}\right)^l
\left({L(\eta)^{1+\nu+\epsilon}\over 1+\nu+\epsilon}\right)
=H_0(\eta)K^{(l,m)}(\eta,\alpha)\ ,
\label{irrelevantse}
\end{eqnarray}
where
\begin{equation}
H_0(\eta)={1\over \nu+1}(1-L(\eta))^{\nu+1}
\label{extraextrao}
\end{equation}
is the ``Born" cross section, and
\begin{equation}
K^{(l,m)}(\eta,\alpha)=\alpha^mc(l,m){l!\over (1+\nu)^l}\sum_{j=0}^l(-1)^j(1+\nu)^j
\ln^j(1-L(\eta))
\label{extraextrat}
\end{equation}
is the ``K-factor" due to the $c(l,m)$ radiative correction in
Eq.~(\ref{explano}).  Near threshold,
\begin{equation}
\lim_{\eta\rightarrow 0}L(\eta)\simeq 1-2\eta \ .
\label{extraextrath}
\end{equation}
While both $H_0(\eta)$ and $H^{(l,m)}(\eta,\alpha)$ vanish as
$\eta\rightarrow 0$, the function $K^{(l,m)}(\eta,\alpha)$
diverges as the $l$-th power of the logarithm.  Relative to the lowest order
term, its contribution can be arbitrarily large in $\eta$ space for
a sufficiently high power $l$, despite the perturbative suppression $\alpha^m$.
The logarithmic terms will have a significant effect on the physical cross
section, obtained as the integral over $\eta$ of Eq.~(\ref{irrelevantse})
convoluted with the parton distributions, especially if the hadronic
center-of-mass energy is such that the support of this integration
emphasizes the threshold region. Under such circumstances we may view
$\eta$ as a second important physical parameter, in addition to $m$, that is
probed phenomenologically. The two-scale nature of the problem is evident
in that we have one hard scale $m$ and a ratio $\eta = (s/4m^2) -1$ whose
important physical domain is near 0.  The perturbative QCD series is required
to have reliable behavior in {\it both} variables, $m$ and $\eta$.

The objective of resummation is to derive formulas, based on the properties
of QCD as a field theory, that provide a summation of the various classes of
monomials in Eq.~(\ref{explano}). In typical resummation methods,
the partial sums of the hard part, Eq.~(\ref{explano}), are replaced with a
resummed hard function that contains the numerically important pieces of the
partial sums, to all orders in pQCD. Most resummations of threshold effects
result in exponentiation of the large logarithmic contributions.  A generic
resummation may result in an expression such as
\begin{equation}
{\cal H}(z,\alpha)\simeq{\rm e}^{E(x_z,\alpha)}\ .
\label{explanth}
\end{equation}
The principal content of resummation resides in
the exponent $E(x_z,\alpha)$, which is typically a function of the QCD
running coupling strength, integrated through intermediate momentum scales.
The particular form of this function is fairly process-independent and
follows from general field-theoretical arguments\cite{ref:stermanold}.
For the present, we concentrate on power-counting, ignoring complications
specific to particular resummation methods.  A large part of the remainder
of the paper is devoted to the complications.

Under specific conditions that we analyze in detail later,
the resummation exponent can be cast into a perturbative form similar to
that of Eq.~(\ref{explano}),
\begin{equation}
E(x_z,\alpha)=\sum_{m=1}^N\alpha^m\sum_{l=0}^{m+1}e(l,m)x_z^l\ .
\label{explant}
\end{equation}
The precise power structure above follows from first principles. For a
resummation method to be realistic in practice, the exponent $E(x_z,\alpha)$
should be calculable in a finite number of steps. One consequence is that the
exponent is not calculable with arbitrary precision, implying, in turn,
a level of uncertainty in some of the coefficients $e(l,m)$, for
specific ranges of $l,m$. The only consistent treatment of this
limitation is to calculate the exponent in enough detail that
the uncalculable coefficients in Eq.~(\ref{explant}) accompany
logarithmic structures $x_z^l$ that are numerically insignificant in the
range where the representation Eq.~(\ref{explant}) makes sense in the
first place. As demonstrated explicitly in the Drell-Yan
process\cite{ref:stermanz}, and as will become apparent below, use of the
two-loop QCD running coupling strength is necessary in the calculation.
Roughly speaking, the representation Eq.~(\ref{explant}) is valid
when $\alpha x_z <1$.  This inequality determines the
limit of calculational accuracy of the exponent.  The upshot
is that the coefficients $e(l,m),\ l\in\{0,m-1\},\ m\ge 3$ are undetermined.

Because the exponent is the central object in resummation, most of the
approximations are effected on that quantity.  By way of definition we say
that we resum {\it leading logarithms} when all $\{e(m+1,m)\}$ are determined,
and {\it non-leading logarithms} when all $\{e(m,m)\}$ are determined
in addition.  Resummation of leading and non-leading logarithms has
been done in detail for the Drell-Yan process \cite{ref:alvero}.
It is important to note, however, that our definition is not
identical to the obvious definition that exists at finite orders.  The latter
is contained in the former.  In the next few paragraphs, we establish the
relationship between the power-counting of logarithms in resummation and the
obvious finite-order definition at the beginning of this section.

Leading-logarithm resummation includes {\it all} leading logarithms at finite
orders. Indeed, suppose we have performed leading-logarithm resummation. Then
\begin{equation}
E(x_z,\alpha)\simeq \sum_{m=1}^N\alpha^me(m+1,m)x_z^{m+1}\ ,
\label{explanf}
\end{equation}
and the kernel of the hard part is
\begin{equation}
{\cal H}(z,\alpha)=\prod_{m=1}^N{\rm e}^{\alpha^me(m+1,m)x_z^{m+1}}\ .
\label{explanfi}
\end{equation}

To compare with finite-order pQCD, we must make
a Taylor expansion of Eq.~(\ref{explanfi}) in $\alpha$. It is
clear that all leading logarithm coefficients in Eq.~(\ref{explano}),
$c(2m,m)$, are obtained from the set $\{e(m+1,m)\}$ of Eq.~(\ref{explanfi})
and, more specifically, from $e(2,1)$ itself. Consider for example the Taylor
expansion of the two first terms of the product of Eq.~(\ref{explanfi}):
\begin{equation}
\sum_{k_1=0}^{\infty}{\alpha^{k_1}x_z^{2k_1}e^{k_1}(2,1)\over k_1!}
\times
\sum_{k_2=0}^{\infty}{\alpha^{2k_2}x_z^{3k_2}e^{k_2}(3,2)\over k_2!}\ .
\label{explans}
\end{equation}
The general monomial of this product is
\begin{equation}
{\alpha^{k_1+2k_2}x_z^{2k_1+3k_2}
e^{k_1}(2,1)e^{k_2}(3,2)\over k_1!k_2!} \ ,
\label{explanse}
\end{equation}
and the only terms that fit the set
described by $c(2m,m)$ are $k_1\in {\bf N}, k_2=0$. Therefore,
$c(2m,m)=e^m(2,1)/m!$, and leading-logarithm resummation includes all
finite-order leading logarithms, along with a specification of the values
of these terms beyond the order in perturbation theory at which they may have
been computed explicitly.  It should be remarked that it includes {\it more}.
The product of Eq.~(\ref{explanfi}) generates upon expansion subleading
logarithms in finite-order pQCD.
For example, $k_1\in {\bf N},k_2=1$ in Eq.~(\ref{explanse}) are
of the first-subleading kind $c(2m-1,m)$ of Eq.~(\ref{explano}).
These are not a closed set, i.e., they account only partially
for all the subleading logarithms resulting from a finite-order calculation.
For example, to account for all $\{c(2m-1,m)\}$, a non-leading exponent
in Eq.~(\ref{explant}) would have to be calculated, i.e., all
coefficients $\{e(m,m)\}$ in Eq.~(\ref{explant}) would in addition
have to be specified.  Upon expansion, such
a resummation of finite-order first subleading logarithms
would produce a series
\begin{equation}
{\cal H}(z,\alpha)=\sum_{m=0}^\infty\alpha^m\sum_{l=m}^{2m}h(l,m)x_z^l\ ,
\label{explanni}
\end{equation}
where each coefficient $h(l,m)$ is a  product of $\{e(l,m)\}$'s.

We remark that Eq.~(\ref{explanni}) does not determine
uniquely all finite-order coefficients $\{c(l\ge m,m)\}$,
{\it for all values of $m$}, but it leaves a subset undetermined. To illustrate
this point, consider the leading uncertainty in the resummation exponent.
Since the one-loop running coupling constant produces all $\{e(m+1,m)\}$
and part of $\{e(m,m)\}$, $m\in {\bf N}$, and the two-loop running
coupling constant produces the remaining pieces of $\{e(m\ge 2,m)\}$,
the leading uncertainty in the exponent is $\delta\alpha^3x_z^2$ in
realistic resummations. This is the lowest-order in $\alpha$,
highest-order in $x_z$ undetermined monomial that comes from expanding the
three-loop running coupling strength in the exponent.

A product of the form
\begin{equation}
\sum_{k_1=0}^{\infty}{\alpha^{k_1}x_z^{2k_1}e^{k_1}(2,1)\over k_1!}
\times
\sum_{k_2=0}^{\infty}{\alpha^{3k_2}x_z^{2k_2}\delta^{k_2}\over k_2!}
\label{explanei}
\end{equation}
would then be present in the resummed hard kernel. The general monomial is
\begin{equation}
{\alpha^{k_1+3k_2}x_z^{2(k_1+k_2)}e^{k_1}(2,1)\delta^{k_2}\over k_1!k_2!} .
\label{irrelevantei}
\end{equation}
We wish to investigate the degree to which the second series
in Eq.~(\ref{explanei})
maximally changes the coefficients $\{h(l,m)\}$ of Eq.~(\ref{explanni}).
Setting $2(k_1+k_2)=2(k_1+3k_2)-n$, we see that we obtain $k_2=n/4$. Hence
the minimum $n=4$, and the minimum $k_2=1$.  Maximum uncertainty arises in the
coefficients
\begin{equation}
h(2m-4,m) .
\label{irrelevantni}
\end{equation}
The coefficients $\{h(l\ge m,m)\}$ are affected {\it for high enough $m$}.
For example at minimum power of $\alpha$, $m=4$, the ``diagonal"
coefficient $h(4,4)$ is affected, at $m=5$ the coefficients
$h(6,5)\ ,h(5,5)$ are affected, etc. However, the leading, first-, second-
and third-subleading logarithms, as defined in finite-order power-counting,
are unaffected for any orders. These finite-order logarithmic structures
are then resummed to all orders, by a resummation of leading and
non-leading logarithms. It is unlikely that the
affected structures are significant numerically in such a
resummation, once we restrict ourselves to the perturbative regime in $x_z$.

We complete this section by returning to the reaction of interest, $t\bar{t}$
production.  At this point, we invoke universality with the Drell-Yan case.
Because the finite-order leading logarithms are identical in the $t\bar{t}$
and $l\bar{l}$ cases, we can resum them in $t\bar{t}$-production with the same
function we use in the Drell-Yan case. According to Ref.~\cite{ref:stermanz},
the structure of the kernel of the resummed hard part in the Drell-Yan case is
\begin{equation}
I(z,\alpha)=\delta(1-z)-\left({{\rm e}^{E(x_z,\alpha)}\over 1-z}
\sum_{j=0}^\infty Q_j(x_z,\alpha)\right)_+\ .
\label{ten}
\end{equation}
For simplicity, we drop the channel indices $ij$;
$x_z\equiv \ln(1/(1-z))$. We defer to sections 3 and 4 an explanation of the
various functions in this equation.  One can therefore write the resummed
partonic cross section as
\begin{equation}
\sigma_{ij}(\eta,m^2)=\int_{1-4(1+\eta)+4\sqrt{1+\eta}}^1
dzI(z,\alpha)\sigma_{ij}^B(\eta,z,m^2)\ .
\label{eleven}
\end{equation}
After integration to get rid of the delta function and plus-distributions,
we find
\begin{equation}
\sigma_{ij}(\eta,m^2)=\int_{1-4(1+\eta)+4\sqrt{1+\eta}}^1
dz{\cal H}(z,\alpha)\sigma'_{ij}(\eta,z,m^2)\ .
\label{thirteen}
\end{equation}
The kernel of the hard part is
\begin{equation}
{\cal H}(z,\alpha)=1+\int_0^{\ln({1\over 1-z})}dx{\rm e}^{E(x,\alpha)}
\sum_{j=0}^\infty Q_j(x,\alpha)\ .
\label{fifteen}
\end{equation}

The kernel in Eq.~(\ref{fifteen}) depends solely on the resummation exponent
$E(x,\alpha)$, either explicitly, or through the functions $Q_j$
which depend exclusively on $E$. This exponent, in turn, depends
on the factorization scheme, i.e., for the process under consideration,
on Eqs.~(\ref{twelve}) and (\ref{twelvep}).
It is to this exponent that we turn our attention in the next section.

\section{The Resummation Exponent}

Invoking universality with the Drell-Yan case, we can express the exponent for
$t\bar{t}$ production using the results of Refs.~\cite{ref:stermanold}.
By the same token, {\it because of the restrictions} on that universality
we may keep only the pieces of the exponent that are universal
in the two cases, i.e., the pieces that reproduce upon expansion the leading
logarithmic structures attributable to initial-state radiation,
exemplified in Eq.~(\ref{twelve}).  The details of the inversion of the
Mellin transform, expressed in Eq.~(\ref{fifteen}), may be found in
Refs.~\cite{ref:stermanz, ref:alvero} and are analysed further in section 4.

In this section, we present the exponent $E(x,\alpha)$ of Eq.~(\ref{fifteen})
in moment space, where the moment $n=\exp(x)$.  For the Drell-Yan
process, the exponent in moment space in the PVR approach may be written in
either the DIS or the $\overline{{\rm MS}}$ factorization scheme
\cite{ref:stermano,ref:lastfour}.  In the DIS scheme,
\begin{equation}
E(x,\alpha)=-\int_P d\zeta{\zeta^{n-1}-1\over 1-\zeta}
\left\{\int_{(1-\zeta)^2}^{(1-\zeta)}
{d\lambda\over \lambda}g_1[\alpha(\lambda m)]+g_2[\alpha((1-\zeta)m)]\right\},
\label{tone}
\end{equation}
and, in the $\overline{{\rm MS}}$ case,
\begin{equation}
E(x,\alpha)=-\int_P d\zeta{\zeta^{n-1}-1\over 1-\zeta}
\int_{(1-\zeta)^2}^1
{d\lambda\over \lambda}g_1[\alpha(\lambda m)].
\label{tonep}
\end{equation}
Here $P$ is a principal-value contour\cite{ref:stermano};
$g_1$ and $ g_2$ are functions of the QCD running coupling
strength $\alpha(\lambda m)$.
It is important to note that Eqs.~(\ref{tone}) and (\ref{tonep}) include
in general all large logarithmic structures in the Drell-Yan case.
This exponent is renormalization-group invariant, by construction.
The logarithmic structures generated by this exponent are recovered
upon expanding the functions $g_i$ as perturbative series with
respect to the running coupling strength and re-expanding the
running coupling strengths in terms of the hard-scale coupling
strength $\alpha(m)$\cite{ref:stermano}.  As we argued in the
general discussion of section 2.2, and as shown explicitly in
\cite{ref:stermano}, a series of truncations can include all large
threshold corrections.

The only necessary
ingredients are the two-loop running coupling strength and a few terms in the
expansions of the $g_i$'s. In general, $g_1$ generates all the leading
threshold corrections and some non-leading ones,  and $g_2$
completes the resummation of numerically important non-leading ones.
Since we base our resummation on universality, which
is valid for leading logarithms only, we disregard $g_2$ for most of the
rest of this paper, and we use the notation $g\equiv g_1$.
We apply the consistency requirement of resumming leading logarithms only by
performing the appropriate truncations and neglecting subleading structures
whenever they appear.  We describe these structures quantitatively in this
section.  We discuss the renormalization-group (RG) properties
of resummation in section 3.4.
The process of resumming leading logarithms only, and the associated
truncations, result in
an approximation of the renormalization-group-invariant exponent of
Eq.~(\ref{tone}) by an exponent that is {\it approximately}
renormalization-group invariant. By
varying the factorization/renormalization scale $\mu$ within a
logical range about its central value $\mu=m$,  we obtain a variation of
the resulting resummed cross section that serves as a quantitative measure
of the effects of logarithmic structures that are not resummable in the
$t\bar{t}$ process.  Variation with $\mu$ is {\it the bulk} of the
theoretical uncertainty.

\subsection{Truncation and perturbative representation of the exponent}

Using the truncations
\begin{equation}
g[\alpha(\lambda m)]=\sum_{j=1}^\infty \alpha^j(\lambda m)g^{(j)}\simeq
\alpha(\lambda m)g^{(1)} ,
\label{ttwo}
\end{equation}
along with
\begin{equation}
\alpha(\lambda m)={\alpha\over 1+\alpha b_2\ln \lambda},\ \ \ \alpha\equiv
\alpha(m) ,
\label{tthree}
\end{equation}
we write the PVR exponent in the $\overline{{\rm MS}}$ scheme as
\begin{equation}
E(x,\alpha)\simeq-\alpha g^{(1)}\int_P d\zeta{\zeta^{n-1}-1\over 1-\zeta}
\int_{(1-\zeta)^2}^1
{d\lambda\over \lambda} {1\over 1+\alpha b_2\ln\lambda}\ .
\label{tfour}
\end{equation}
This integral can be evaluated exactly\cite{ref:stermano}.  The result is
\begin{equation}
E(x,\alpha)=- {g^{(1)}\over b_2}{1\over t}
\sum_{m=1}^\infty {(1-n)_m\over m!m^2}{\cal E}(mt) ,
\label{tfive}
\end{equation}
where
\begin{equation}
t\equiv {1\over 2\alpha b_2},\ \ (1-n)_m={\Gamma(1-n+m)\over \Gamma (1-n)},\ \
{\cal E}(y)=y{\rm e}^{-y}{\rm Ei}(y),
\label{tsix}
\end{equation}
and the exponential integral is defined by the principal value
\begin{equation}
{\rm Ei}(y)\equiv {\cal P}\int_{-\infty}^ydx{\rm e}^x/x\ .
\label{irrelevantten}
\end{equation}

Equation~(\ref{tfour}) or, equivalently, Eq.~(\ref{tfive}) has a perturbative
asymptotic representation\cite{ref:stermano}
\begin{equation}
E(x,\alpha)\simeq E(x,\alpha,N(t))=g^{(1)}\sum_{\rho=1}^{N(t)+1}\alpha^\rho
\sum_{j=0}^{\rho+1}s_{j,\rho}x^j\ .
\label{teight}
\end{equation}
This representation is valid in the moment-space interval
\begin{equation}
1<x\equiv \ln n< t .
\label{tseven}
\end{equation}
The coefficients are
\begin{equation}
s_{j,\rho}=-b_2^{\rho-1}(-1)^{\rho+j}2^\rho c_{\rho+1-j}(\rho-1)!/j! ,
\label{tnine}
\end{equation}
and the constants $c_k$ are obtained from the expansion
$\Gamma(1+z)=\sum_{k=0}^\infty c_k z^k$, where $\Gamma$ is the Euler gamma
function.

The number of perturbative terms $N(t)$ in Eq.~(\ref{teight}) is
obtained by optimizing the asymptotic approximation
\begin{equation}
\bigg|E(x,\alpha)-E(x,\alpha,N(t))\bigg|={\rm minimum} .
\label{tten}
\end{equation}
Using Eqs.~(\ref{tfive}), (\ref{teight}), and (\ref{tnine}), we rewrite
Eq.~(\ref{tten}) as
\begin{equation}
\Bigg|\sum_{m=1}^n{(1-n)_m\over m!m^2}{\cal E}(mt)-\sum_{\rho=0}^{N(t)}
{(-1)^\rho\rho!\over t^\rho}
\sum_{j=0}^{\rho+2} {(-1)^{j-1}c_{\rho+2-j}\over j!}x^j\Bigg|={\rm minimum}\ .
\label{televen}
\end{equation}
Equation~(\ref{televen}) denotes an approximation between
two functions of the moment $n$, and it determines the optimum number of
perturbative terms $N$ as a function of the parameter $t$.
Indeed, as shown numerically below, within the interval of Eq.~(\ref{tseven}),
the optimization of Eq.~(\ref{televen}) has a solution that depends
on $t$ {\it only}, $N=N(t)$. It also can be shown\cite{ref:alvero}  that
in the complementary interval
\begin{equation}
t<x\equiv \ln n<\infty ,
\label{ttwelve}
\end{equation}
the approximation Eq.~(\ref{tnine}) breaks down for any integer $N$
that is a function of $t$ {\it only}. For very large $n$ within
the interval of Eq.~(\ref{ttwelve}) the asymptotic approximation
\begin{equation}
E(x,\alpha)\simeq -{g^{(1)}\over b_2}
\left\{t\left({x\over t}-1\right)\ln\left({x\over t}-1\right)-x\right\}
\label{tthirteen}
\end{equation}
holds, and it is clearly a non-perturbative one.

Throughout this paper, we use the two-loop formula for the fixed coupling
strength
\begin{equation}
\alpha(m)\equiv {\alpha_s(m)\over \pi}={1\over b_2\ln(m^2/\Lambda^2)}
-{b_3\over b_2^3}{\ln(\ln(m^2/\Lambda^2))\over \ln^2(m^2/\Lambda^2)} ,
\label{tfourteen}
\end{equation}
with
\begin{equation}
b_2=(11C_A-2n_f)/12,\ b_3=(34C_A^2-(10C_A+6C_F)n_f)/48 ,
\label{tfifteen}
\end{equation}
and number of flavors $n_f=5$.  We set $\Lambda=0.158$ GeV (the ${\rm CTEQ3M}$
value\cite{ref:cteq}).

In Fig.~1 we illustrate the validity of the asymptotic approximation
for a value of t corresponding to $m=175$ GeV.  In
Fig.~1a we show how $N(t)$ is determined from Eq.~(\ref{televen})
for a fixed $t$ and selected parametric values of $n$.
The plot shows that optimization
works perfectly, and it demonstrates the typical breakdown of the asymptotic
approximation as $N$ increases beyond $N(t)$. This rise
is the exponential rise of the infrared renormalons,
the $\rho!$ growth in the second term of Eq.~(\ref{televen}).
As long as $n$ is in the interval of Eq.~(\ref{tseven}),
all the members of the family in $n$ are optimized
at the same $N(t)$, showing that the optimum number of
perturbative terms is a function of $t$ only.
In Fig.~1b we plot the function $N(t)$ for a range of $t$ relevant to
$t\bar{t}$ production.  An excellent numerical approximation is provided
by the fit
\begin{equation}
N(t)\simeq [t-3/2] ,
\label{tsixteen}
\end{equation}
where the integer part is defined as the closest integer from either direction.
(It is amusing that this fit suggests a lower limit for the hard scale $m$.
A perturbative series is an improvement in accuracy if $N(t)\ge 1$ which,
from the above fit, implies $t-3/2\ge 0.5$.  Using for simplicity
a one-loop fit of $\alpha(m)$ and $\Lambda$, we deduce that the hard scale
$m/\Lambda\ge {\rm e}^2\simeq 8$, well within our expectations.)

Equations~(\ref{tseven}) and (\ref{televen}) suggest a
perturbative behavior for the exponent in moment space
{\it independently of} the color factors which reside in $g^{(1)}$.
In particular, the range of validity of the perturbative
expression for the exponent, $x/t< 1$, is obtained by {\it direct
comparison} with the exact principal-value definition, as shown
in Fig.~2. (Equations~(\ref{ttwelve}) and (\ref{tthirteen})
are also suggestive, but they constitute an asymptotic limit,
valid for $x$ larger than $t$.)
This range of validity has the consequence that terms in the exponent of the
form $\alpha^k\ln^kn$ are of order unity, and terms with fewer powers
of logarithms, $\alpha^k\ln^{k-m}n$, are negligible.
This explains why resummation is completed in a finite number of steps in
the Drell-Yan process, as discussed earlier.
The same is true here, but, in addition, we discard monomials
$\alpha^k\ln^kn$ in the exponent because of the restricted
universality between the $t\bar{t}$ and $l\bar{l}$ processes.

The exponent we use in the rest of the paper is
the truncation
\begin{equation}
E(x,\alpha,N)=g^{(1)}\sum_{\rho=1}^{N(t)+1}\alpha^\rho s_\rho x^{\rho+1} ,
\label{tseventeen}
\end{equation}
with the coefficients
\begin{equation}
s_\rho\equiv s_{\rho+1,\rho}=b_2^{\rho-1}2^\rho/\rho(\rho+1)\ .
\label{teighteen}
\end{equation}
We note in passing that the leading logarithm truncation of the exponent,
Eqs.~(\ref{tseventeen}) and (\ref{teighteen}), forms a convergent
series,
\begin{equation}
E(x,\alpha,N)\simeq {g^{(1)}\over b_2}x\left\{1+\left({1\over 2\alpha b_2 x}-1\right)
\ln(1-2\alpha b_2 x)\right\} ,
\label{convergent}
\end{equation}
as long as $x<1/(2\alpha b_2)\equiv t$.
This, of course, is not true for the exponent of Eq.~(\ref{teight})
that contains the full subleading logarithmic structures in the
Drell-Yan process.
We use the convergent version of the exponent Eq.~(\ref{convergent})
in special cases to arrive at simplified expressions for the
perturbative regime in momentum space.

There are equivalent expressions appropriate for the DIS factorization scheme.
Equation~(\ref{tfour}) becomes
\begin{eqnarray}
& &E(x,\alpha)\simeq -\alpha g^{(1)}\int_Pd\zeta{\zeta^{n-1}-1\over 1-\zeta}
\int_{(1-\zeta)^2}^{1-\zeta}{d\lambda\over \lambda}
{1\over 1+\alpha b_2\ln\lambda}\nonumber \\
& &=
-{g^{(1)}\over b_2}\Biggl\{{1\over t}\sum_{m=1}^\infty{(1-n)_m\over m!m^2}
{\cal E}(mt)-{1\over 2t}
\sum_{m=1}^\infty {(1-n)_m\over m!m^2}{\cal E}(2mt)\Biggr\} \ .
\label{tninteen}
\end{eqnarray}
The perturbative approximation, subject to the truncation requirements
discussed previously, is
\begin{equation}
E(x,\alpha,N)=g^{(1)}\sum_{\rho=1}^{N(t)+1}\alpha^\rho s_\rho x^{\rho+1}-
g^{(1)}\sum_{\rho=1}^{N(2t)+1}\left({\alpha\over 2}\right)^\rho s_\rho
x^{\rho+1}\ ,
\label{ttwenty}
\end{equation}
or, in convergent form,
\begin{equation}
E(x,\alpha,N)\simeq {g^{(1)}\over b_2}x\left\{\left({1\over 2\alpha b_2 x}-1\right)
\ln(1-2\alpha b_2x)-\left({1\over \alpha b_2 x}-1\right)\ln(1-\alpha b_2 x)\right\}
\ .
\label{convergentdis}
\end{equation}

The convergent exponents, Eqs.~(\ref{convergent}) and (\ref{convergentdis}),
suggest the perturbative interval of Eq. (\ref{tseven}), independently of PVR.
On the other hand, beyond the end point $x=1/(2\alpha b_2)$ these convergent
expressions are ill-defined because the leading term has a branch-point
singularity.

\subsection{Regularization-independence of the perturbative exponent}

It is valuable to stress that we can derive the perturbative expressions,
Eqs.~(\ref{teight}), (\ref{tseven}), (\ref{tnine}), and (\ref{tsixteen}),
without the PVR prescription, although with less certitude.  The analysis of
this subsection is presented in order to show that our final perturbative
results do not depend on the specific manner that the infra-red region is
regularized in the PVR approach.

We begin with the {\it unregularized} form of Eq.~(\ref{tfour}), i.e.,
with the integral over $x$ done on the real axis:
\begin{equation}
E_0(x,\alpha)=-\alpha g^{(1)}\int_0^1dx{x^{n-1}-1\over 1-x}\int_{(1-x)^2}^1
{d\lambda\over \lambda}{1\over 1+\alpha b_2\ln\lambda}\ .
\label{teighteenp}
\end{equation}
We expand the inner integrand as a Taylor series around $\alpha$.
As we will see, part the problem now transforms into ignorance
of the asymptotic  properties of this expansion. Writing
\begin{equation}
\int_{(1-x)^2}^1{d\lambda\over \lambda}{1\over 1+\alpha b_2\ln\lambda}=
2\sum_{\rho=1}{(-1)^\rho\over \rho t^{\rho-1}}\ln^\rho (1-x)\ ,
\label{teighteenpp}
\end{equation}
we deduce
\begin{equation}
E_0(x,\alpha)=-2\alpha g^{(1)}\sum_{\rho=1}{(-1)^\rho\over \rho t^{\rho-1}}
\int_0^1{dy\over y}\biggl[(1-y)^{n-1}-1\biggr]\ln^\rho y\ .
\label{teighteenppp}
\end{equation}
The upper limit of the summation in Eq.~(\ref{teighteenppp}) is left
undetermined, because, as is made evident below, the series represented
is only formal, i.e., not convergent.  Lack of convergence is associated
with the Landau pole exhibited
by the original integral, Eq.~(\ref{teighteenp}).  Using the identity
\begin{equation}
\ln^\rho y=\lim_{\eta\rightarrow 0}\left({\partial\over \partial\eta}\right)^\rho
y^\eta\ ,
\label{irrelevantel}
\end{equation}
and the Stirling approximation for the beta function $B(n,\eta)$,
\begin{equation}
\lim_{\eta\rightarrow 0}\left({\partial\over  \partial \eta}\right)^\rho
\left(B(n,\eta)-{1\over \eta}\right)=\rho!\sum_{j=0}^{\rho+1}{(-1)^j\over j!}
c_{\rho+1-j}\ln^\rho n\ ,
\label{irrelevanttwe}
\end{equation}
we find
\begin{equation}
E_0(x,\alpha)=g^{(1)}\sum_{\rho=1}\alpha^\rho\sum_{j=0}^{\rho+1}s_{j,\rho}x^j\ ,
\ x\equiv\ln n,
\label{teighteenpppp}
\end{equation}
with the coefficients of Eq.~(\ref{tnine}).  This expression for $E_0(x,\alpha)$
 is the same as Eq.~(\ref{teight}), the only and major difference being
that we do not know the asymptotic properties of this series in
the full range of moments $n$. The added information is precisely what is
furnished by PVR, as we saw earlier. Both the function $N(t)$ {\it and}
the range of validity of the perturbative expression, $1<x\equiv \ln n< t$,
are provided by the principal-value prescription.

To be more explicit, we examine Eq.~(\ref{teighteenpppp}) in some detail.
Because the coefficients $s_{j,\rho}$ grow factorially,
the series does not represent a convergent infinite series for a fixed
$t$ and $n$.  The factorial growth
is precisely the infrared renormalon induced by the existence
of the Landau-pole singularity of the original integral.
Without this factorial, as in the truncated
expression Eq.~(\ref{tseventeen}), the resulting infinite series
is convergent for a fixed $n$, but it diverges nevertheless at the threshold
$n\rightarrow \infty$ due to the powers of the logarithm. Since $E$ is
the exponent of the cross section, the resulting singularities
in the cross section would be essential singularities.
In fact, in Eqs.~(\ref{teight}) and (\ref{tnine}) there is a trade-off
between factorial growth and powers of logarithms:  greater factorial growth
is accompanied by fewer powers of moment logarithms (and hence of momentum
logarithms in the cross section).

We conclude that the nature of the series in Eq.~(\ref{teighteenpppp}) is
asymptotic, and we rewrite Eq.~(\ref{teighteenpppp}) as
\begin{equation}
E_0(x,\alpha)\simeq E(x,\alpha,N)=g^{(1)}\sum_{\rho=1}^{N}
\alpha^\rho\sum_{j=0}^{\rho+1}s_{j,\rho}x^j\ ,
\ x\equiv\ln n\ .
\label{teighteenppppp}
\end{equation}
Because the original integral is unregulated,
the properties and range of validity of this asymptotic series are not obvious.
In PVR, regularization is incorporated, and, since there are
no undetermined extra scales, such as the introduction of IR cutoffs, the
asymptotic properties are determinable fully, as shown in section 3.1

For a fixed $t$ and $n$, one may use
the monotonicity behavior of the corresponding partial
sums to try to determine an
upper limit for the number of terms in Eq.~(\ref{teighteenppppp}).
This procedure is illustrated in Fig.~3a for $m=175$ GeV. We note that
beyond a certain range of $N$, the exponent increases factorially,
a demonstration of both the asymptotic nature of the series and
of the effect of the IR renormalons.  A range of
optimum $N$ can be determined where the growth of the sum reaches a
plateau, before the factorial growth sets in at large $N$.
The exponent is fairly flat in
this region so the indeterminacy of the optimum $N$
is insignificant numerically.  The plateau is centered
around $N_{opt}\simeq $ 6 to 7 for a wide range of moments,
in agreement with the results of the previous subsection.
One can make these statements quantitative by defining the slope
\begin{equation}
S(x,\alpha,N)\equiv {\partial E(x,\alpha,N)\over \partial N}=
E(x,\alpha, N+1)-E(x,\alpha,N)\ .
\label{slopedef}
\end{equation}
For a range of $n$ where there is a plateau, the optimum $N$ can be determined
by the equation
\begin{equation}
N_{opt}:\ S(x,\alpha,N)\bigg|_{x,\alpha=fixed}={\rm minimum}\ .
\label{noptimum}
\end{equation}
In Fig.~3b, we observe that the optimization of the
perturbative exponent with this method gives the same results
as the principal-value method, i.e.,
\begin{equation}
N_{opt}-1\simeq N(t)=[t-3/2] .
\label{equalns}
\end{equation}

A second issue is whether one could determine the same range of
validity, Eq.~(\ref{tseven}), of the representation Eq.~(\ref{teighteenppppp})
without PVR. This is possible if we impose the {\it supplementary}
requirement that this representation results in a value of $N_{opt}$ that
depends {\it only on the hard scale} $m$, and not on the moment $n$.
In Fig.~3b we show $N_{opt}$, the minimum of the slope $S(n,N)$ for various
values of $n$ at $m=175$ GeV.  We observe that $N_{opt}$ is constant in
approximately the same range of $n$ as in Eq.~(\ref{tseven}).
When $n$ is increased beyond that
range, the corresponding $N_{opt}$ becomes a function of $n$.
As  $n$ is increased,
the plateau created by the corresponding values of the exponent
starts shrinking, and the renormalons move to the left of the figure.
 For values of $n$ close to the boundary of Eq.~(\ref{tseven}), the slope in
Eq.~(\ref{slopedef}) becomes substantial.

The two requirements, a plateau in $E(x,\alpha,N)$ and a value of $N_{opt}$
that depends only on the hard scale, allow one to make statements
concerning the perturbative representation of the exponent that are similar
to those the PVR approach, albeit with less certitude.
The perturbative representation of the exponent,
Eqs.~(\ref{teight}), (\ref{tseven}), (\ref{tnine}), and (\ref{tsixteen}), is
obtained in straightforward fashion in PVR, but the analysis of this subsection
shows that the result is approximately independent of
the regularization PVR imposes.  The result can be recovered by a study of
the factorial growth of the partial sums and the reasonable physical requirement
that the plateau of stability of the exponent depend on the
hard scale of the problem only.

\subsection{Perturbative regime of the hard part in moment space}

We have established a perturbative representation
for the exponent in moment space, with or without principal-value
regularization.  The conclusions are similar.
The exponent has a perturbative representation in the
range
\begin{equation}
1<x\equiv \ln n<t\equiv {1\over 2\alpha b_2}
\label{exprange}
\end{equation}
independent of the constant $g^{(1)}$, i.e., of the channel-dependent
color factors. This is a {\it maximum} range allowed
for the perturbative cross section in moment space.
One should distinguish, however, between the range in moment space
where the exponent has a perturbative representation, and
the range where the cross section itself is perturbative.
The distinction arises because the cross section, which is proportional
in moment space to the {\it exponential} of $E(\ln n,\alpha,N)$,
is much more sensitive to variations in $n$.  The cross section also depends
exponentially on the color factors $g^{(1)}$, which can be much larger
than unity, while the perturbative representation of the exponent
in the regime of Eq.~(\ref{exprange}) is independent of the color factors.

We address the question of the perturbative regime in momentum space in
section 4, where we present the inverse Mellin transform that provides
the cross section. However, one can furnish a heuristic argument
in moment space as well. Again, this argument is independent of
the regularization prescription of the original integral representation
of the exponent, once we work with its asymptotic perturbative
approximation, Eqs.~(\ref{teight}) and (\ref{tseventeen}).

The idea is to regard the plateau of Fig.~3a as a region
of {\it perturbative stability} in both of the variables it depends on,
namely $N$ and $n$. Denote the kernel of the hard part in moment space by
\begin{equation}
{\tilde I}(n,\alpha,N)={\rm e}^{E(\ln n,\alpha, N)}\ .
\label{hardmoment}
\end{equation}
Then, this kernel is perturbatively stable in an interval of $n$ such
that variations around the center of the plateau $N(t)$, which
does not depend on $n$ in the perturbative regime, provide
${\cal O}(\alpha_s)$ variations for ${\tilde I}(n,\alpha,N)$, i.e.,
numerically negligible contributions that are not enhanced by
threshold effects. This statement is made quantitative by the
requirement
\begin{equation}
\delta_N{\tilde I}(n,\alpha,N(t))={\rm e}^{E(\ln n,\alpha,N(t))}
\delta_NE(\ln n,\alpha,N(t))={\rm e}^{E(\ln n,\alpha,N(t))}S(\ln n,\alpha,N(t))
\le \alpha_s\ ,
\label{pertconstraintn}
\end{equation}
where the slope $S$ is defined in Eq.~(\ref{slopedef}).
Another way of reaching the condition of Eq.~(\ref{pertconstraintn})
begins with the order-by-order perturbative hard part written as
\begin{equation}
{\tilde I}^{[k]}(n,\alpha)=\sum_{m=0}^k\alpha^ms_m(n)
+\sum_{m=1}^k\alpha^mr_m(n)\ .
\label{hardseries}
\end{equation}
The first sum in Eq.~(\ref{hardseries}) contains singular functions of $n$
that produce large threshold enhancements and must
be resummed, while  the second sum contains regular functions of $n$,
whose numerical effects are of ${\cal O}(\alpha_s)$.
The first sum is replaced by our resummed ${\tilde I}(n,\alpha,N(t))$,
and the condition in Eq.~(\ref{pertconstraintn}) is equivalent to
a plateau of perturbative stability such that variations
of the resummed terms around the plateau belong to the second, regular, sum
of Eq.~(\ref{hardseries}):
\begin{eqnarray}
& &{\rm e}^{E(\ln n,\alpha,N(t))+\delta_NE(\ln n,\alpha,N(t))}\simeq
{\rm e}^{E(\ln n,\alpha,N(t))}
+{\rm e}^{E(\ln n,\alpha,N(t))}S(n,\alpha,N(t))\nonumber \\
& &\simeq {\rm e}^{E(\ln n,\alpha,N(t))}+\sum_{m=1}^{l}\alpha^mr'_m(n) ,
\label{variation}
\end{eqnarray}
so that $\delta_N{\tilde I}(n,\alpha,N(t))\simeq{\cal O}(\alpha_s)$.
This is the same criterion as Eq.~(\ref{pertconstraintn}).

The regular sums in Eqs.~(\ref{hardseries}) and (\ref{variation}),
have been calculated only to first order in $t\bar{t}$ production.
This is the reason
we consider them at most equal to $\alpha_s$, rather than assuming
definite numerical coefficients, large or small, unsupported by
exact calculations.  This statement
is corroborated by the exact one-loop calculation of the $t\bar{t}$ cross
section, and by the one- and two-loop calculations of the Drell-Yan
cross section. It is in general assumed in all resummation procedures.

\subsection{Renormalization-group properties of the perturbative exponent}

We have studied the exponent under the simplification
that the renormalization/factorization scale, the ``hard scale", is fixed at
$\mu=m$.  In this subsection, we discuss the dependence on $\mu$
of the resummation exponent in moment space.  We
work entirely in the $\overline{{\rm MS}}$ scheme.
Since RG invariance is one of the ingredients of resummation,
we expect that the full Drell-Yan exponent, containing the full scale
dependence, is exactly RG invariant.
It can be seen by inspection that Eq.~(\ref{tonep}) is scale-invariant:
\begin{equation}
E(x,\alpha)=E(x,\alpha(\mu),m/\mu)=
-\int_P
d\zeta{\zeta^{n-1}-1\over 1-\zeta}
\int_{(1-\zeta)^2m^2/\mu^2}^{m^2/\mu^2}
{d\lambda\over \lambda} g_1[\alpha(\lambda\mu)] .
\label{muonep}
\end{equation}

Here we ask to what degree the truncations we impose, starting
from Eq.~(\ref{tfour}),  affect this invariance.
We begin with the explicitly $\mu$-dependent equivalent
of  Eq.~(\ref{tfour}), read from Eq.~(\ref{muonep}).
The corresponding expression is\cite{ref:lastfour}
\begin{equation}
E(x,\alpha(\mu),m/\mu)\simeq-\alpha(\mu) g^{(1)}\int_P
d\zeta{\zeta^{n-1}-1\over 1-\zeta}
\int_{(1-\zeta)^2m^2/\mu^2}^{m^2/\mu^2}
{d\lambda\over \lambda} {1\over 1+\alpha(\mu) b_2\ln\lambda}\ ,
\label{muone}
\end{equation}
where the dependence on $\mu$ is both implicit (in $\alpha(\mu)$) and explicit.
Working along the lines of section 3.2, we may show that
\begin{eqnarray}
& &E(x,\alpha(\mu),m/\mu)\simeq E(x,\alpha(\mu),m/\mu,N(\mu))\nonumber \\
& &=
-g^{(1)}\sum_{\rho=1}^{N(\mu)+1}\alpha^\rho(\mu){(-1)^\rho 2^\rho b_2^{\rho-1}
\over \rho}\sum_{k=1}^\rho{\rho!\over (\rho-k)!}\ln^{\rho-k}(m/\mu)
\sum_{j=0}^{k+1}{(-1)^j\over j!}c_{k+1-j}x^j\ .
\label{mutwo}
\end{eqnarray}
We use the approximation $N(\mu)\simeq N(t(\mu))$.
This equation makes clear that {\it explicit} $\mu$-variation (i.e. $k < \rho$)
is equivalent to inclusion of {\it non-leading} logarithms $x^j, j < \rho + 1$.
In this sense, uncertainty expressed through $\mu$-variation
overlaps with uncertainty expressed through inclusion of non-universal
logarithms.  In the triple sum of Eq.~(\ref{mutwo}),
the $k=\rho$ term is the dominant term
numerically in the exponent.  This term represents the implicit
$\mu$-dependence, Eq.~(\ref{teight}).
All other $\ln(m/\mu)$ terms are non-leading
in $x = \ln n$.  The linear term in $\ln(m/\mu)$, obtained for $k=\rho-1$,
has a universal $\ln n$ substructure. It is the one containing equal
powers of $\alpha(\mu)$ and $\ln n$.  It comes from the
same diagrams as the leading $\ln n$ structures at $\mu=m$ and
amounts to exponentiation of the explicit ${\cal O}(\alpha_s^3)$
$\mu$-variation.

Our $\mu$-dependent exponent for $t\bar{t}$ production can therefore
be written
\begin{equation}
E(x,\alpha(\mu),m/\mu,N(\mu))=g^{(1)}\Biggl\{\sum_{\rho=1}^{N(\mu)+1}
\alpha^\rho(\mu){b_2^{\rho-1}2^\rho\over \rho(\rho+1)}x^{\rho+1}-\ln(m/\mu)
\sum_{\rho=1}^{N(\mu)+1}\alpha^\rho(\mu){b_2^{\rho-1}2^\rho\over \rho}
x^\rho\Biggr\}\ .
\label{muthree}
\end{equation}
We call attention to the sign structure of Eq.~(\ref{muthree}).  For a fixed
moment and scales $\mu$ decreasing from $m$, the implicit growth of the first
sum due to the increase of
$\alpha(\mu)$ is cancelled by the minus sign of the explicit dependence
of the second sum, and vice-versa.
On the other hand, in the truncation from Eq.~(\ref{mutwo}) to
Eq.~(\ref{muthree}), non-universal subleading logarithmic structures are
discarded.  Therefore we expect only approximate scale invariance of
Eq.~(\ref{muthree}).

In Fig.~4 we plot the normalized exponent, in each of the
truncated versions of Eqs.~(\ref{muone}), (\ref{mutwo}), and (\ref{muthree}),
versus a wide variation of $\mu\in \{100, 300\}$ GeV,
and for various perturbative values of the moment $n$.
We see that both the PV Drell-Yan exponent Eq.~(\ref{muone}) and
its perturbative Drell-Yan version Eq.~(\ref{mutwo}) are straight lines,
practically coincident.  The universal exponent Eq.~(\ref{muthree})
is a gently increasing function of $\mu$, and it lies somewhat below the
Drell-Yan exponent.  The increase
creates a partial compensation of the explicit $\mu$-dependence
of the Born cross sections, more so than if the full Drell-Yan exponent
were used. This difference stands to reason, since no implicit $\mu$-dependence
exists in the Drell-Yan process at the Born level.
These $\mu$-dependence properties conform to the intuition that our resummed
cross section should show less variation with $\mu$ than the
next-to-leading order counterpart.  We call attention to
Fig.~4b, where we show the effects of $\mu$-dependence
as a function of the number of terms in the partial sums. For the Drell-Yan
exponent, the addition of perturbative terms {\it up to the optimum number}
$N(t(\mu))$ makes the exponent progressively scale-invariant.

Smooth behavior under scale transformations is not a property of the
resummation of Ref.~\cite{ref:laeneno}.  If an IR cutoff is used,
conventional RG-scale invariance looses much of its meaning
since an additional and arbitrary scale is introduced.
We return to this issue in section 5 when we discuss the physical cross
section and compare scale variations in our approach with
the corresponding ones of Ref.~\cite{ref:laeneno}.

\section{Leading and subleading logarithmic structures}

In this section we describe the resummed partonic cross section
in momentum space, derive a perturbative approximation in a specific
region of momentum space, and discuss in general how to include
leading logarithmic structures consistently in the resummation.
We also demonstrate how
the constant $g^{(1)}$ is determined from the one-loop calculation, and
we discuss the issue of universality of the leading logarithms.
In particular, we demonstrate that retention of subleading pieces
in the exponent, of a structure similar to the Drell-Yan case, would not
account correctly for the subleading structure at the one-loop level
in top quark production.  We justify the truncations we use consistently
with universality of initial-state radiation.

The general form of the PVR cross section in the $\overline{{\rm MS}}$
scheme is
\begin{equation}
\sigma_{ij}^{PV}(\eta,m^2)=\int_{1-4(1+\eta)+4\sqrt{1+\eta}}^1
dz{\cal H}(z,\alpha)\sigma'_{ij}(\eta,z,m^2)\ ,
\label{thone}
\end{equation}
where the kernel of the hard part is
\begin{equation}
{\cal H}(z,\alpha)=1+\int_0^{x_z}dx{\rm e}^{E(x,\alpha)}
\sum_{j=0}^\infty Q_j(x,\alpha) ,
\label{thtwo}
\end{equation}
and
\begin{equation}
x_z\equiv \ln\left({1\over 1-z}\right)\ .
\label{thtwop}
\end{equation}
The functions $Q_j(x,\alpha)$ in Eq.~(\ref{thtwo}) appear
during the inversion of the Mellin transform
$n\leftrightarrow z$\cite{ref:stermanz}.  They
are obtained from the generating function
\begin{eqnarray}
& &{\cal Q}[P_1,P_2,...P_{N+1}]\equiv {\rm Re}\Biggl\{{1\over i\pi}
\sum_{m_1,m_2,...m_{N+1}=0}^\infty {P_1^{m_1}P_2^{m_2}...P_{N+1}^{m_{N+1}}
\over m_1!m_2!...m_{N+1}!}\nonumber \\
& &\times {\rm lim}_{\epsilon\rightarrow 0}
\left({\partial\over \partial\epsilon}\right)^{m_1+2m_2+...(N+1)m_{N+1}}
{\rm e}^{i\pi\epsilon}\Gamma(1+\epsilon)\Biggr\}
\label{ththree}
\end{eqnarray}
through the identification
\begin{equation}
{\cal Q}[P_1,P_2,...P_{N+1}]\equiv \sum_{j=0}^\infty Q_j\ .
\label{thfour}
\end{equation}
This notation is slightly different from that of Ref.~\cite{ref:stermanz}.
In particular, in the present notation the real part is extracted explicitly
from the generating function.  In Eqs.~(\ref{ththree}) and (\ref{thfour}),
\begin{equation}
P_k=P_k(x,\alpha)\equiv {\partial^k E(x,\alpha)\over k!\partial x^k} .
\label{thfive}
\end{equation}
The function $Q_j=Q_j(x,\alpha)$ is defined as the set of terms in
Eq.~(\ref{thtwo}) that contribute $j$ more powers of $\alpha$ than of $x$.
The structures $Q_j$ resum the $j$-th subleading
logarithms in the physical cross section through the index identity
\begin{equation}
j=\sum_{k=2} m_k(k-1)
\label{irrelevantthi}
\end{equation}
that connects Eqs.~(\ref{ththree}) and (\ref{thfour}).

\subsection{The leading resummed perturbative cross section and its
range of validity}

For best accuracy, in a process like the Drell-Yan process,
the expression for the hard part should include enough of the set of $Q_j$'s
to reproduce the resummable finite-order subleading logarithms up to
two-loops. On the other hand, owing
to the constrained universality that characterizes the $t\bar{t}$
cross section, we should include only terms that do not create
non-universal subleading logarithms.  In addition, we must discuss
the range in momentum space where our perturbative resummation
is valid.

We turn first to the issue of the perturbative regime in momentum
space. Specification  of this region follows from general expressions for the
inversion of the Mellin transform and the meaning of the successive terms
in this inversion, once their perturbative approximations are used.
The kernel of the hard part is provided in Eq.~(\ref{thtwo}),
and the functions $Q_0$ and $Q_1$, obtained
from Eqs.~(\ref{ththree}) and (\ref{thfive}), are
\begin{equation}
Q_0={1\over \pi}\sin(\pi P_1)\Gamma(1+P_1)
\label{thseven}
\end{equation}
and
\begin{eqnarray}
& &Q_1\simeq 2\Gamma(1+P_1)P_2\cos(\pi P_1)\Psi(1+P_1)\ .
\label{thtwentytwo}
\end{eqnarray}
The functions $\Psi\equiv\Psi^{(0)}$ and $\Psi^{(k)}$ are the usual Polygamma
functions.  For simplicity we include in the expression for
$Q_1$ only terms that generate corrections
starting at ${\cal O}(\alpha)$.   According to
the general discussion at the beginning this section,
$Q_1$ contributes one less power of $x$ than of $\alpha$
in the integrand of Eq.~(\ref{thtwo}), and it is formally subleading
relative to the contribution of $Q_0$. Nevertheless, from
Eqs.~(\ref{thseven}) and (\ref{thtwentytwo}), we see that this
suppression is not true
for values of $x$ such that $P_1(x,\alpha)\simeq 1$.  When
$\sin(\pi P_1)\simeq 0$, the dominance of
$Q_0$ over $Q_1$ is destroyed.  For values of $x$ such that
$P_1>1$, $Q_0$ and $Q_1$ are out of phase, with amplitudes
that are not constrained, and the perturbative dominance of
$Q_0$ over $Q_1$ is again vitiated.

We conclude that the perturbative region in momentum space is defined by
the inequality constraint
\begin{equation}
P_1(x_z,\alpha)\le 1\ .
\label{theight}
\end{equation}
In our discussion in moment space in section 3, we saw that the perturbative
approximation
for the exponent is valid in the interval $1<\ln n\le t=1/2\alpha b_2$, i.e.,
where terms containing equal powers of $\ln n$ and $\alpha$ are at most
${\cal O}(1)$.  Our
Eq.~(\ref{theight}) translates this condition consistently into momentum space.
In Fig.~5 the upper two curves show the range in $\ln n = x$ where
the constraint
of Eq.~(\ref{theight}) is satisfied.  The lower three curves in Fig.~5 pertain
to the corresponding constraint for the hard part in moment space,
Eq.~(\ref{pertconstraintn}). We see that the two criteria are in
good agreement: the value of $\ln n = x$ at which
Eq.~(\ref{theight}) is satisfied
agrees fairly well with the value of $n$ at which Eq.~(\ref{pertconstraintn}) is
also.

We take up next the issue of universal logarithmic
structures in our resummation.  Power-counting, discussed generically in
section 2.2, must be altered when applied to the kernel of the integral in
Eq.~(\ref{thtwo})
because the extra integration supplies one more power of the logarithm.
We begin with the following expression for the kernel of the hard part:
\begin{equation}
{\cal H}(z,\alpha)\simeq 1+\int_0^{x_z}dx{\rm e}^{E(x,\alpha)}
Q_0(x,\alpha)\ ,
\label{thsix}
\end{equation}
and we rewrite
\begin{equation}
Q_0(x,\alpha)=Q_0(x,\alpha)-P_1(x,\alpha)+P_1(x,\alpha)\equiv
\Delta(x,\alpha)+P_1(x,\alpha)\ .
\label{definediff}
\end{equation}
Using Eq.~(\ref{tseventeen}), we can express
\begin{equation}
{\rm e}^{E(x,\alpha)}=\sum_{m=0}^{\infty}\alpha^m\sum_{l=m+1}^{2m}
\epsilon(l,m)x^l ,
\label{exponentaylor}
\end{equation}
and, from the definitions of $Q_0$ and $P_1$, we find the expansion
\begin{equation}
\Delta(x,\alpha)=\sum_{m=2}^\infty\alpha^m\delta_mx^m .
\label{deltataylor}
\end{equation}
Therefore,
\begin{equation}
\int_0^{x_z}dx{\rm e}^{E(x,\alpha)}\Delta(x,\alpha)=
\sum_{m=0}^\infty \alpha^{m+2}\sum_{k=0}^m\sum_{l=k+1}^{2k}
{\epsilon(l,k)\delta_{m+2-k}\over 3+m+l-k}
x_z^{3+m+l-k}\ .
\label{totaltaylor}
\end{equation}
For a fixed $m$, the maximum monomial in powers of $x_z$ is
$\alpha^{2+m}x_z^{2m+3}$.  In the notation of section 2.2,
the integral in Eq.~(\ref{totaltaylor}) contributes {\it at most}
to the first-subleading
logarithms $c(2m-1,m)$. Since we intend to resum leading logarithms only,
we ignore the contribution from the integral Eq.~(\ref{totaltaylor}).
For the same reason, we discard all $Q_j$'s but $Q_0$ in
Eq.~(\ref{thtwo}).

Our main result for the perturbative resummed
partonic cross section, denoted by $\sigma_{ij}^{pert}$, is therefore
\begin{eqnarray}
& &\sigma_{ij}^{pert}(\eta,m^2)=\int_{1-4(1+\eta)+4\sqrt{1+\eta}}^{z_0}
dz\biggl[1+
\int_0^{x_z}dx{\rm e}^{E(x,\alpha)}
P_1(x,\alpha)
\biggr]\sigma'_{ij}(\eta,z,m^2)\nonumber \\
& &=\int_{1-4(1+\eta)+4\sqrt{1+\eta}}^{z_0}
dz{\rm e}^{E(x_z,\alpha)}\sigma'_{ij}(\eta,z,m^2)\ .
\label{thten}
\end{eqnarray}
The upper limit of integration $z_0\equiv z_0(\alpha)$ is determined from the
equation for the perturbative regime
\begin{equation}
{1\over 1-z_0}={\rm e}^{x_0}\ ,\ \ \ P_1(x_0,\alpha)=1\ .
\label{ponemeretime}
\end{equation}

The preceding expressions in this subsection are valid in either factorization
scheme. The only distinction is the value of the multiplicative constant
$g^{(1)}$ in Eq.~(\ref{teighteen}).
This constant can be determined from a comparison of the ${\cal O}(\alpha)$
expansion of the resummed partonic cross section
with the finite-order calculation of Refs.~\cite{ref:dawson,ref:vneerven}.

It is useful for purposes of comparison to expand
the cross section to ${\cal O}(\alpha^2)$.
In the $\overline{{\rm MS}}$ factorization scheme, using
Eqs.~(\ref{tseventeen}) and (\ref{teighteen}), we obtain
\begin{equation}
E^{[2]}(x,\alpha)=\alpha g^{(1)}x^2+2\alpha^2g^{(1)}b_2x^3/3\ .
\label{theleven}
\end{equation}
Substituting Eq.~(\ref{theleven}) into Eq.~(\ref{thten}), we find that
the corresponding finite-order part of the resummed cross section is
\begin{eqnarray}
& &\sigma_{ij}^{[2]}(\eta,m^2)\bigg|_{pert}=
\int_{1-4(1+\eta)+4\sqrt{1+\eta}}^1dz\Biggl\{1+\alpha g^{(1)}
\ln^2\biggl({1\over 1-z}\biggr)\nonumber \\
& &+\alpha^2\Biggl[{(g^{(1)})^2\over 2}
\ln^4\biggl({1\over 1-z}\biggr)+
{2\over 3}g^{(1)}b_2\ln^3\biggl({1\over 1-z}\biggr)\Biggr]\Biggr\}\sigma'_{ij}(\eta,z,
m^2)\ .
\label{thtwelve}
\end{eqnarray}

To determine $g^{(1)}$, one can
use the corresponding ${\cal O}(\alpha)$ expression for
the cross section from Ref.~\cite{ref:laeneno},
or derive an asymptotic formula near threshold
($\eta\rightarrow 0$) for the ${\cal O}(\alpha)$ piece of
the integration Eq.~(\ref{thtwelve}) and compare with the corresponding
explicit expression of Ref.~\cite{ref:dawson}. The former is simpler for the
present purposes.  In section 4.2, we derive the asymptotic threshold
formula to demonstrate universality issues.
The result is
\begin{equation}
g^{(1)}_{\overline{{\rm MS}}}=2C_{ij},\ \ C_{q\bar{q}}=C_F,\ \ C_{gg}=C_A\ ,
\label{ththirteen}
\end{equation}
the same as in the Drell-Yan case.
Our Eq.~(\ref{thtwelve}), including the ${\cal O}(\alpha^2)$ term, is
identical to that of Ref.~\cite{ref:laeneno}.

Similarly, in the DIS factorization scheme, using Eqs.~(\ref{ttwenty}) and
(\ref{teighteen}), we derive
\begin{equation}
E^{[2]}(x,\alpha)=\alpha g^{(1)}x^2/2+\alpha^2g^{(1)}b_2x^3/2\ .
\label{thfourteen}
\end{equation}
Hence the cross section to ${\cal O}(\alpha^2)$ becomes
\begin{eqnarray}
& &\sigma_{ij}^{[2]}(\eta,m^2)\bigg|_{pert}=
\int_{1-4(1+\eta)+4\sqrt{1+\eta}}^1dz\Biggl\{1+\alpha {g^{(1)}\over 2}
\ln^2\biggl({1\over 1-z}\biggr)\nonumber \\
& &+\alpha^2\Biggl[{1\over 2}\biggl({g^{(1)}\over 2}\biggr)^2
\ln^4\biggl({1\over 1-z}\biggr)+
{1\over 2}g^{(1)}b_2\ln^3\biggl({1\over 1-z}\biggr)\Biggr]\Biggr\}\sigma'_{ij}(\eta,z,
m^2)\ .
\label{thfifteen}
\end{eqnarray}
By comparison with the corresponding ${\cal O}(\alpha)$ piece of
Ref.~\cite{ref:laeneno}, we obtain
\begin{equation}
g^{(1)}_{{\rm DIS}}=2C_{ij}=g^{(1)}_{\overline{{\rm MS}}}\ ,
\label{thsixteen}
\end{equation}
again in agreement with the Drell-Yan case.
As before, Eq.~(\ref{thfifteen}), including the ${\cal O}(\alpha^2)$-term,
is identical to that of Ref.~\cite{ref:laeneno}.

We close this subsection with simplified analytical expressions
for the perturbative regime of the cross section, Eq.~(\ref{theight}).
For this purpose, we may use the convergent expressions,
Eqs.~(\ref{convergent}) and (\ref{convergentdis}). In the
$\overline {\rm MS}$ factorization scheme, we obtain
\begin{equation}
P_1(x_z,\alpha)\simeq -{g^{(1)}\over b_2}\ln(1-2\alpha b_2x_z)\le 1 ,
\label{convponems}
\end{equation}
and in the DIS scheme,
\begin{equation}
P_1(x_z,\alpha)\simeq
-{g^{(1)}\over b_2}\ln(1-2\alpha b_2x_z)
+{g^{(1)}\over b_2}\ln(1-\alpha b_2x_z)
\le 1 .
\label{convponedis}
\end{equation}
To obtain transparent analytical expressions, we specialize here to
the one-loop approximation
\begin{equation}
2\alpha b_2=\ln^{-1}(m/\Lambda) .
\label{simple}
\end{equation}
We obtain
\begin{equation}
1-z\ge \left({\Lambda\over m}\right)^{1-{\rm exp}(-b_2/g^{(1)})}\ \ \
{\overline {\rm MS}},
\label{highertwistms}
\end{equation}
\begin{equation}
1-z\ge \left({\Lambda\over m}\right)^{1-{
{\rm exp}(-b_2/g^{(1)})/2
\over
1-{\rm exp}(-b_2/g^{(1)})/2}}\ \ \
{\rm DIS}.
\label{highertwistdis}
\end{equation}
We observe that the non-perturbative regime is suppressed
mainly by $\Lambda/m$ to the first power, with a
correction that depends on the color factors of the
partonic production process.  For $m=175$ GeV,
\begin{equation}
\Lambda/m\simeq 10^{-3}.
\label{firstpower}
\end{equation}
Because $b_2$ is positive ($b_2\simeq 2$), the non-perturbative
regime is an increasing function of the color factors.
If the partonic cross sections had no color-factor enhancements,
$g^{(1)}\le 1$, Eqs.~(\ref{highertwistms}) and (\ref{highertwistdis})
would be in close agreement with the perturbative regime of the
exponent in moment space $n\le m/\Lambda$, Eq.~(\ref{tseven}), with the
direct substitution $1/n\leftrightarrow 1-z$.
For example, for $g^{(1)}=1$,
\begin{equation}
1-z\ge 2\times 10^{-3}\ \ \ {\overline{\rm MS}}\ {\rm and}\
1-z\ge 1.5\times 10^{-3}\ \ \ {\rm DIS}\ ,
\label{htwistsch}
\end{equation}
in good agreement with Eq.~(\ref{firstpower}).
In reality, $g^{(1)}=8/3$ in the $q\bar{q}$ channel, therefore
\begin{equation}
1-z\ge 2\times 10^{-2}\ \ \ {\overline{\rm MS}}\ {\rm and}\
1-z\ge 8\times 10^{-3}\ \ \ {\rm DIS}\ .
\label{htwistschemes}
\end{equation}
For the $gg$ channel, $g^{(1)}=6$, and we find
\begin{equation}
1-z\ge 1\times 10^{-1}\ \ \ {\overline{\rm MS}}\ {\rm and}\
1-z\ge 5\times 10^{-2}\ \ \ {\rm DIS}\ .
\label{htwistschemess}
\end{equation}
These regions are narrower than the estimate based
on the perturbative properties of the exponent only,
especially for the $gg$ channel.  The soft gluon region is probed more
deeply in the DIS scheme than in the
${\overline{\rm MS}}$ scheme, closer to $\Lambda/m$.
We discuss these properties numerically in section 5.

\subsection{Non-universal subleading logarithms}

In this subsection, we address universality of the logarithmic structures
in the threshold region more explicitly.
It is sufficient to examine the issue in the $\overline{{\rm MS}}$
scheme.  For completeness, and to compare with the analytical
results of Ref.~\cite{ref:dawson}, we derive the near-threshold asymptotic
properties of the ${\cal O}(\alpha)$ term of Eq.~(\ref{thtwelve}).
We demonstrate that if we were to keep the subleading logarithm in the
Drell-Yan exponent, Eqs.~(\ref{teight}) and (\ref{tnine}), the resulting
term in the ${\cal O}(\alpha)$ expansion of Eq.~(\ref{thtwo}) would not be
the same as the one obtained in the explicit next-to-leading order calculation
of Ref.~\cite{ref:dawson}. This is a demonstration of the consistency
requirements of the series of truncations that resulted in
Eqs.~(\ref{tseventeen}), (\ref{teighteen}), and (\ref{thten}).

In the ${\overline{\rm MS}}$ scheme,  Eq.~(\ref{teight}),
the full ${\cal O}(\alpha)$ Drell-Yan exponent is
\begin{equation}
E^{[1]}(x,\alpha)=\alpha g^{(1)}(s_{2,1}x^2
+s_{1,1}x+s_{0,1})\ ,
\label{thtwentyone}
\end{equation}
where the coefficients are found in Eq.~(\ref{tnine}).  Suppose
we allow an undetermined linear term, $\alpha \kappa x$,
that would represent the deviation of this exponent from the Drell-Yan form.
We discard the constant term in this exercise, because it does not contribute
logarithmic terms to the cross section at first order.
Substituting from Eq.~(\ref{tnine})
we have
\begin{equation}
E^{[1]}(x,\alpha) \simeq
\alpha (g^{(1)}x^2+2g^{(1)}\gamma x +\kappa x) ,
\label{thtwentyonep}
\end{equation}
where $\gamma$ is the Euler-Mascheroni constant.
The kernel of the hard part, containing subleading logarithmic structures, is
given generally by Eq.~(\ref{thtwo}). To account for subleading logarithms
with full accuracy in the Drell-Yan case, we must keep $Q_0$ and $Q_1$,
Eqs.~(\ref{thseven}) and (\ref{thtwentytwo}).
Expanding to ${\cal O}(\alpha)$ and
substituting into Eq.~(\ref{thone}), we obtain
\begin{equation}
\sigma_{ij}^{(1)}(\eta, m^2)=\alpha
\int_{1-4(1+\eta)+4\sqrt{1+\eta}}^1
dz\left[g^{(1)}\ln^2\left({1\over 1-z}\right)+\kappa\ln\left({1\over 1-z}\right)
\right]\sigma'_{ij}(\eta,z,m^2)\ .
\label{thtwentythree}
\end{equation}
The piece of the Drell-Yan expression linear in the logarithm has dropped out.
In other words, the exact evaluation of the cross section
of Eq.~(\ref{thtwentythree}) with the Drell-Yan resummation expression
gives the same answer as the truncated expression, Eq.~(\ref{thtwelve}).
This is something well-known in the Drell-Yan case and is
expressed by absence of the function $g_2$ in
the $\overline{{\rm MS}}$ scheme, Eq.~(\ref{tonep}).

The integral representation of the partonic cross section does not contain
the subleading logarithms of the Drell-Yan case.
This is not true for its asymptotic evaluation near threshold, as we
shall see below, owing to the form of the Born cross section.
We focus on the $q\bar{q}$ channel only.
From Eqs.~(\ref{fifteenp}) and (\ref{thtwentythree}), we obtain
\begin{eqnarray}
& &\sigma_{q\bar{q}}^{(1)}(\eta,m^2)=
{8\over 9}\pi^3\alpha^3{\tau\over s}
\int_{1-4(1+\eta)+4\sqrt{1+\eta}}^1dz
[1-(1-z)\tau]\times\nonumber \\
& &\left[\sqrt{[1-(1-z)\tau]^2-4\tau}+{2\tau\over \sqrt{[1-(1-z)\tau]^2-
4\tau}}\right]\times\nonumber \\
& &\left[g^{(1)}\ln^2\left({1\over 1-z}\right)+\kappa\ln
\left({1\over 1-z}\right)\right] ,
\label{thtwentyfour}
\end{eqnarray}
where $\tau\equiv m^2/s$.  Introducing the variables
$x=[1-(1-z)\tau]^2-4\tau$ and\cite{ref:dawson} $\beta^2=1-4\tau$, we derive
\begin{eqnarray}
& &\sigma_{q\bar{q}}^{(1)}={4\over 9s}\alpha_s^3\int_0^{\beta^2}dx\biggl[x^{1/2}
+(1-\beta^2)x^{-1/2}/2\biggr]\times\nonumber \\
& &\left[g^{(1)}
\ln^2\left({4\over 1-\beta^2}\bigl[1-\sqrt{1-(\beta^2-x)}\bigr]\right)
+\kappa\ln\left({4\over 1-\beta^2}\bigl[1-\sqrt{1-(\beta^2-x)}\bigr]\right)\right]
\ .
\label{thtwentyfive}
\end{eqnarray}
The threshold is exhibited at $\beta\rightarrow 0$.
Expanding
\begin{equation}
1-\sqrt{1-(\beta^2-x)}\simeq (\beta^2-x)/2\ ,
\label{irrelevantfou}
\end{equation}
and using
\begin{equation}
\ln^k x=\lim_{\epsilon\rightarrow 0}\left({\partial\over \partial \epsilon}
\right)^k
x^\epsilon\ ,
\label{irrelevantfif}
\end{equation}
we find
\begin{eqnarray}
& &\sigma_{q\bar{q}}^{(1)}\simeq{4\over 9}\alpha_s^3{\beta^3\over m^2}\Biggl\{g^{(1)}
\left({\partial\over\partial\epsilon}\right)^2\left({2\beta^2\over 1-\beta^2}\right)^\epsilon
\left[B(3/2,1+\epsilon)+{1-\beta^2\over 2\beta^2}B(1/2,1+\epsilon)\right]\nonumber \\
& &+\kappa {\partial\over \partial\epsilon}
\left({2\beta^2\over 1-\beta^2}\right)^\epsilon\left[B(3/2,1+\epsilon)+
{1-\beta^2\over 2\beta^2}B(1/2,1+\epsilon)\right]\Biggr\}\ .
\label{thtwentysix}
\end{eqnarray}
After computing the derivatives, we obtain
\begin{eqnarray}
& &\sigma_{q\bar{q}}^{(1)}\simeq{4\over 9}\alpha_s^3{\tau\over m^2}\beta^3
\Biggl[ g^{(1)}
\Biggl\{\ln^2\left({2\beta^2\over 1-\beta^2}\right)\left(B(3/2,1)+
{1-\beta^2\over 2\beta^2}B(1/2,1)\right)\nonumber \\
& &+2\ln\left({2\beta^2\over 1-\beta^2}\right)\left(B(3/2,1)(\Psi(1)-\Psi(5/2))
+{1-\beta^2\over 2\beta^2}B(1/2,1)(\Psi(1)-\Psi(3/2))\right)\nonumber \\
& &+B(3/2,1)\left((\Psi(1)-\Psi(5/2))^2+\Psi'(1)-\Psi'(5/2)\right)\nonumber \\
& &+{1-\beta^2\over 2\beta^2}B(1/2,1)\left((\Psi(1)-\Psi(3/2))^2+
\Psi'(1)-\Psi'(3/2)\right)\Biggr\}\nonumber \\
& &+\kappa\Biggl\{\ln\left({2\beta^2\over 1-\beta^2}\right)\left(B(3/2,1)+
{1-\beta^2\over 2\beta^2}B(1/2,1)\right)\nonumber \\
& &+B(3/2)(\Psi(1)-\Psi(5/2))+{1-\beta^2\over 2\beta^2}B(1/2,1)(\Psi(1)-
\Psi(3/2))\Biggr\}\Biggr] \ .
\label{thtwentyseven}
\end{eqnarray}
Discarding the non-logarithmic pieces and using the
relations $\Psi(1)=-\gamma$,
$\Psi(3/2)=-\gamma-\ln 4+1$, we observe that all the coefficients of the
logarithms can be rational. Introducing\cite{ref:dawson} $\rho=4\tau$,
we end with
\begin{equation}
\sigma_{q\bar{q}}^{(1)}\simeq{\alpha_s^3\over m^2}{\beta\rho\over 9}
\left\{g^{(1)}\ln^2(8\beta^2)-[2g^{(1)}-\kappa]\ln(8\beta^2)\right\}\ .
\label{thtwentyeight}
\end{equation}
Our Eq.~(\ref{thtwentyeight}) may be compared with the result of
Ref.~\cite{ref:dawson} in the $\overline{{\rm MS}}$
scheme, denoted $\alpha_s^3f_{q\bar{q}}^{(1)}$:
\begin{equation}
\alpha_s^3f_{q\bar{q}}^{(1)}\simeq{\alpha_s^3\over m^2}{\beta\rho\over 9}
\left\{{8\over 3}\ln^2(8\beta^2)-{41\over 6}\ln(8\beta^2)\right\}\ .
\label{thtwentynine}
\end{equation}
We conclude that, for the leading logarithm,
\begin{equation}
g^{(1)}=8/3=2C_F\ ,
\label{ththirty}
\end{equation}
the same value found in Eq.~(\ref{ththirteen}).  However, for the
next-to-leading linear logarithm,
\begin{equation}
2g^{(1)}=16/3\ne 41/6\ .
\label{ththirtyone}
\end{equation}
A non-zero value for $\kappa$ would be required.  This exercise shows
that universality between the Drell-Yan case and $t\bar{t}$
production is restricted to the leading logarithms only, as must be
the resummation of these structures.  The subleading logarithms
in the full exponent of Eq.~(\ref{teight}) are uncertain.

The analysis of this subsection
justifies our truncated expressions, and it motivates several equivalent ways
to define a quantitative measure of the overall theoretical uncertainty.
The most straightforward is to vary the renormalization/factorization (hard)
scale within a reasonable range and to use the variation
of the cross section as a measure of the non-universal subleading
structures.  It is important to recall, however,
that truncation of the resummation of leading logarithms at a given
{\it higher order} produces subleading logarithms at that order, as the
${\cal O}(\alpha^2)$ terms in Eqs.~(\ref{thtwelve}) and (\ref{thfifteen}) show.
We will assume that these particular subleading logarithms are approximately
universal because they come from the evolution properties of the resummed
universal leading logarithms of lower orders. This is true for the
$\overline{{\rm MS}}$ Drell-Yan cross section at two
loops\cite{ref:drellyantwo}.

To summarize, the resummation exponent in the perturbative regime
in the $\overline{{\rm MS}}$ factorization scheme is
\begin{eqnarray}
& &E(x,\alpha(\mu),m/\mu,N(\mu))=\nonumber \\
& &2C_{ij}\Biggl\{\sum_{\rho=1}^{N(t(\mu))+1}
\alpha^\rho(\mu){b_2^{\rho-1}2^\rho\over \rho(\rho+1)}x^{\rho+1}-\ln(m/\mu)
\sum_{\rho=1}^{N(t(\mu))+1}\alpha^\rho(\mu){b_2^{\rho-1}2^\rho\over \rho}
x^\rho\Biggr\}\ .
\label{thseventeen}
\end{eqnarray}
In the DIS factorization scheme, it is
\begin{eqnarray}
& &E(x,\alpha(\mu),\mu,N(t(\mu)))=\nonumber \\
& &2C_{ij}\Biggl\{\sum_{\rho=1}^{N(t(\mu))+1}
\alpha^\rho(\mu){b_2^{\rho-1}2^\rho\over \rho(\rho+1)}x^{\rho+1}-\ln(m/\mu)
\sum_{\rho=1}^{N(t(\mu))+1}\alpha^\rho(\mu){b_2^{\rho-1}2^\rho\over \rho}
x^\rho\Biggr\}\nonumber \\
& &-2C_{ij}\Biggl\{\sum_{\rho=1}^{N(2t(\mu))+1}
\alpha^\rho(\mu){b_2^{\rho-1}\over \rho(\rho+1)}x^{\rho+1}-\ln(m/\mu)
\sum_{\rho=1}^{N(2t(\mu))+1}\alpha^\rho(\mu){b_2^{\rho-1}\over \rho}
x^\rho\Biggr\}\ .
\label{theighteen}
\end{eqnarray}
The upper integer in the two sums is $N(t(\mu))=[t(\mu)-3/2]$, and
$t(\mu)\equiv 1/(2\alpha(\mu) b_2)$.
The perturbative resummed cross section in the corresponding scheme
is obtained from the general expression Eq.~(\ref{thtwo}), within
the region defined
by Eq.~(\ref{theight}). This restriction can be incorporated if we
define $z_0$ through
\begin{equation}
P_1\left(\ln\left({1\over 1-z_0(\mu,m/\mu)}\right),\alpha(\mu),m/\mu,N(t(\mu))\right)=1\ .
\label{thnineteen}
\end{equation}
Imposing the constraint that only leading logarithms are resummed, we arrive at
our {\it perturbative} resummed cross section
\begin{equation}
\sigma_{ij}^{pert}(\eta, m^2,\mu^2)=
\int_{1-4(1+\eta)+4\sqrt{1+\eta}}^{z_0(\mu,m/\mu)}
dz{\rm e}^{E(x_z,\alpha(\mu),m/\mu,N(t(\mu)))}\sigma'_{ij}(\eta,z,m^2)\ .
\label{thtwenty}
\end{equation}
The derivatives of the Born cross sections, $\sigma'_{ij}(\eta,z,m^2)$, are
found in Eqs.~(\ref{fifteenp}) and (\ref{fifteenpp}).

We use the formulas above to calculate our predictions for
the perturbative resummed cross section.
The rest of the phase space in Eq.~(\ref{thtwenty}), from $z = z_0$ to the
threshold $z = 1$, is non-perturbative, and we declare ignorance about
 the cross section in that region. (In section
6 we engage in some speculations on non-perturbative physics.)
A measure of the dominance of the universal logarithmic structures
and their resummation in the perturbative regime in our approach
is provided by the change in the cross section when the
renormalization/factorization scale is varied
within a reasonable range around the ``central" value $\mu=m$.
This variation measures the importance of
subleading logarithmic structures that are non-universal and
non-resummable in this approach.
We use variation with $\mu$ as a working hypothesis for the bulk
of the perturbative uncertainty\cite{ref:edpapero}.

\section{The resummed cross section}

In this section we present analytical and numerical results for
the resummed cross sections.  Section 5.1 is devoted to the partonic
cross sections as functions of the variable $\eta$, while in section 5.2
we present the physical cross sections obtained after convolution of the
partonic cross sections with parton densities.  We furnish details on the
technicalities involved, both kinematical and numerical.  In section 5.3
we make comparisons with the predictions of Laenen, Smith and van
Neerven\cite{ref:laeneno, ref:laenent}, denoted LSvN.

\subsection{Partonic cross sections}

The resummed partonic cross section has the form
\begin{equation}
\sigma_{ij}^{pert}(\eta, m^2,\mu^2)=
\int_{1-4(1+\eta)+4\sqrt{1+\eta}}^{z_0(\mu,m/\mu)}
dz{\rm e}^{E(x_z,\alpha(\mu),m/\mu,N(t(\mu)))}\sigma'_{ij}(\eta,z,m^2)\ .
\label{fone}
\end{equation}
The upper limit of integration, $z_0(\mu,m/\mu)$, is provided by
Eq.~(\ref{thnineteen}).
For physical but relatively large values of $\eta$, the lower limit
of integration may become negative, unlike the Drell-Yan process in which
$0<z<1$.  In this situation, far away from threshold, resummation of
initial-state gluon radiation is irrelevant, and we do not perform resummation
outside the range $0<z<1$.  Finding the roots of the lower limit of
integration in Eq.~(\ref{fone}) we see that this equation is unconstrained for
\begin{equation}
\eta\le (1+\sqrt{2})^2/4-1=0.4571068\ .
\label{ftwo}
\end{equation}
Equation~(\ref{ftwo}) defines the region in phase space in which threshold
effects are important.  Above this value of $\eta$, Eq.~(\ref{fone}) should
be constrained so that it includes {\it only} the phase space where gluon
radiation is produced near threshold.

In order to achieve the best accuracy available we wish to include in
our predictions as much as is known theoretically.  The exact next-to-leading
order (one-loop) cross section is known\cite{ref:dawson,ref:vneerven},
but the full two-loop
calculation does not exist.  Thus the best we can do at present is
to include the full content of the one-loop partonic cross section along with
our resummation of the leading logarithms to all orders.  In so doing,
we include both
non-universal one-loop subleading logarithms and constants.  This procedure is
in common with previous resummations of this process\cite{ref:laeneno}
and the Drell-Yan process\cite{ref:alvero}.
Our ``final" resummed partonic cross section can therefore be
written\cite{ref:edpapero}
\begin{equation}
\sigma^{pert}_{ij}(\eta, m^2,\mu^2)=\sigma^{pert}_{ij}(\eta,m^2,\mu^2)-
\sigma^{(0+1)}_{ij}(\eta,m^2,\mu^2)\Bigg|_{pert}+\sigma^{(0+1)}_{ij}(\eta,m^2,\mu^2)\ .
\label{fthree}
\end{equation}
The second term is the part of the partonic cross section up to one-loop that is
included in the resummation, while the last term is the exact one-loop
cross section\cite{ref:dawson,ref:vneerven}.

For numerical purposes it is best to convert the ``improper"
integrations induced in Eq.~(\ref{fone}) by the form
of the Born cross sections, $\sigma_{ij}'(\eta, z, m^2)$
(see Eqs.~(\ref{fifteenp}) and (\ref{fifteenpp})),
to integrations without a numerical (but analytically integrable) singularity.
This is achieved by the transformation
\begin{equation}
x=\sqrt{1-{4\tau\over (1-\tau(1-z))^2}}\ .
\label{ffour}
\end{equation}
In addition, it is easier to parametrize the perturbative regime, $z_0$,
by the moment variable $n_0$:
\begin{equation}
{1\over 1-z_0}=n_0,\ \ n_0={\rm e}^{x_0} ,
\label{ffourp}
\end{equation}
where $x_0$ is the first root of the equation
\begin{equation}
P_1(x_0(\mu,m/\mu),\alpha(\mu),m/\mu,N(t(\mu)))=1\ .
\label{ffive}
\end{equation}
Taking into account the constraint on $\eta$, Eq.~(\ref{ftwo}),
we can write the various pieces of Eq.~(\ref{fthree}) as follows:
\begin{equation}
\sigma^{pert}_{ij}(\eta,m^2,\mu^2)=\int_{L(\tau)}^{U(\tau,n_0)}
dx{\rm e}^{E(y(x),\alpha(\mu),m/\mu,
N(t(\mu)))}{\tilde \sigma}_{ij}(x)\ ,
\label{fsix}
\end{equation}
where
\begin{equation}
U(\tau,n_0(\mu,m/\mu))\equiv\sqrt{1-{4\tau\over (1-\tau/n_0(\mu,m/\mu))^2}}\ ,
\label{fsixp}
\end{equation}
\begin{equation}
L(\tau)\equiv \Theta\left({1\over (1+\sqrt{2})^2}-\tau\right)\times
\sqrt{1-{4\tau\over (1-\tau)^2}}\ ,
\label{fseven}
\end{equation}
and
\begin{equation}
y(x)=\ln\left({\tau\sqrt{1-x^2}\over \sqrt{1-x^2}-2\sqrt{\tau}}\right)\ .
\label{fsevenp}
\end{equation}
The second term in Eq.~(\ref{fthree}) is obtained
from Eq.~(\ref{fsix}) by expanding the exponential up to ${\cal O}(\alpha)$.
The exponents for the two factorization schemes are found in
Eqs.~(\ref{thseventeen}) and (\ref{theighteen}), while
the transformed functions ${\tilde\sigma}_{ij}(x)$ are
\begin{equation}
{\tilde\sigma}_{q\bar{q}}(x)={32\over 3}\pi^3\alpha^2{\tau^{3/2}\over s}
\left({1+x^2\over (1-x^2)^{5/2}}\right) \ ,
\label{feight}
\end{equation}
and
\begin{eqnarray}
& &{\tilde\sigma}_{gg}(x)={1\over 2}\pi^3\alpha^2{\tau^{1/2}\over s}
{1\over (1-x^2)^{3/2}}\Biggl[-\biggl({4\over 3}
x^2+5\tau+{12\tau\over 1-x^2}\biggr)x^2\nonumber \\
& &+\biggl(2x^4-{8\over 3}x^2+3\tau x^2
+2-3\tau\biggr)x\ln\biggl({1+x\over 1-x}\biggr)\nonumber \\
& &+\biggl(-{8\over 3}x^4-4x^2+\tau x^2+{4\over 3}-3\tau\biggr)\Biggr].
\label{fnine}
\end{eqnarray}

In Fig.~6 we show the perturbative boundary $n_0(\mu,m/\mu)$, obtained
from Eqs.~(\ref{ffourp}) and (\ref{ffive}), as a function of the hard scale
$\mu$, for $m=175$ GeV.  The scale $\mu$ is an artifact of
perturbation theory, and $\mu$-variation is associated with truncation of the
perturbative expansion. In a resummation such as ours, all significant
perturbative knowledge of threshold effects is exhausted (all large
perturbative threshold corrections are included), and the perturbative regime
we calculate should also be insensitive to artifacts such as $\mu$.
Therefore, one expects the perturbative boundary to be independent of the
hard scale $\mu$ and to depend only on the physical scale characterizing the
threshold, $m$.  We observe that the full Drell-Yan exponent
produces a function $P_1$ that is almost exactly hard-scale invariant,
establishing that our resummation conforms to this intuition.
On the other hand, the truncation to universal logarithmic structures that we
use for our predictions shows some scale dependence.

In Fig.~7 we show the resummed partonic cross sections $\sigma_{ij}^{pert}$
as a function of $\eta$, for $\mu=m$ and  $m=175$ GeV,
for both production channels in the $\overline{{\rm MS}}$ factorization
scheme.  We choose a logarithmic scale in $\eta$ to expand the threshold
region.  We also show the lowest order and next-to-leading order
counterparts.  The three curves differ substantially in the partonic threshold
region $\eta<1$, with the final resummed curve exceeding
the other two.  Above $\eta\simeq 1$, our resummed cross sections are
essentially identical to the next-to-leading order cross sections,
as is to be expected since the near-threshold enhancements that concern us
in this paper are not relevant at large $\eta$.  In both the $q {\bar q}$ and
the $g g$ channels, we note that the size of the
${\cal O}(\alpha_s^3)$ term exceeds that of the ${\cal O}(\alpha_s^2 )$ term
for $\eta \simeq$ 0.1, and the ratio grows as $\eta$ decreases.  This
behavior is contrary to the
notion underlying perturbation theory, that successive terms in the
perturbation series should be smaller, and is cited in our Introduction as the
motivation for resummation at small $\eta$.

It is useful to translate our definition of the perturbative regime directly
into a statement about the perturbative region in $\eta$.
Our perturbative resummation probes the threshold down to the point
\begin{equation}
\eta\ge \eta_0 ={1\over 2 n_0} ,
\label{ften}
\end{equation}
where $n_0$ is calculated from Eqs.~(\ref{ffourp}) and (\ref{ffive}).
Below this value, perturbation theory, resummed or otherwise,
is not to be trusted. For our central predictions we choose
to accept the exact next-to-leading order results throughout the  phase space,
but the non-perturbative region is a source of non-perturbative uncertainty,
subject to model-building.  At $m=175$ GeV, our resummed cross sections
become identical to the next-to-leading order cross sections below
$\eta \geq$ 0.008 for the $q{\bar q}$ channel and $\eta \geq$ 0.05 for the
$gg$ channel, a consequence of our decision to restrict resummation to the
perturbative domain.  The difference reflects the larger color factor in
the $gg$ case.

We return to possible non-perturbative contributions in section 6.

\subsection{The physical cross sections}

We use $S$ to denote the square of the hadronic center-of-mass energy.  Once
the partonic cross sections are calculated, the physical cross section
for each production channel is obtained through the factorization
theorem.
\begin{equation}
\sigma_{ij}(S,m^2)={4m^2\over S}\int_0^{{S\over 4m^2}-1}d\eta\Phi_{ij}\biggl[
{4m^2\over S}(1+\eta),\mu^2\biggr]\sigma_{ij}(\eta,m^2,\mu^2) .
\label{feleven}
\end{equation}
The parton flux is a convolution of parton distributions
\begin{equation}
\Phi_{ij}[y,\mu^2]=\int_y^1{dx\over x}f_{i/h_1}(x,\mu^2)f_{j/h_2}(y/x,\mu^2)\ .
\label{ftwelve}
\end{equation}
We use ${\rm CTEQ}3$ parton distributions\cite{ref:cteq} in the
appropriate factorization scheme.
The total physical cross section is obtained after incoherent addition of the
contributions from  the the $q\bar{q}$ and $gg$ production channels.
We ignore the small contribution from the $qg$ channel.

A quantity of phenomenological interest is the differential cross section
\begin{equation}
{d\sigma_{ij}(S,m^2,\eta)\over d\eta}={4m^2\over S}\Phi_{ij}\biggl[
{4m^2\over S}(1+\eta),\mu^2\biggr]\sigma_{ij}(\eta,m^2,\mu^2)\ .
\label{fthirteen}
\end{equation}
The differential distribution is a RG-invariant quantity, and it is perhaps
measurable.  Its integral over $\eta$ is, of course, the total cross section.
In Fig.~8 we plot these distributions for the two production
channels for $m=175$ GeV, ${\sqrt {S}}=1.8$ TeV and $\mu=m$.  Convolution
with the parton flux enhances the relative importance of the region of small
$\eta$.  We observe that, at the energy of the Tevatron, resummation is
significant for the $q\bar{q}$ channel and less so for the
$gg$ channel.  Figure 8 is useful for back-of-the-envelope estimates of
possible contributions from the non-perturbative regime, discussed in section 6.

We show the total $t\bar{t}$-production cross section for various energies
in Fig.~9, and in Table~1 we provide numerical values.
The central value of our predictions is obtained with the
choice $\mu/m=1$, and the lower and upper limits are our estimate of the
perturbative uncertainty.  These upper and lower values are  the maximum
and minimum of the cross section in the range of the hard scale
$\mu/m\in\{0.5,2\}$.
For the range of top quark mass shown, the minimum occurs at $\mu/m=2$, while
the maximum occurs at $\mu/m\simeq 0.7$, as is also shown in Fig.~9b.
Our prediction of Fig.~9a is in reasonable agreement with the
published data\cite{ref:cdfdz}.
The $\mu$-variation of our resummed cross section, shown in Fig.~9b is smaller
than that of the next-to-leading order cross section, as expected for an
all-orders resummation. (Owing to a computer compiler error, the
variation with $\mu$ of the next-to-leading order results in the first paper of
Ref.~\cite{ref:edpapero} is incorrect.  Our overall resummed predictions
are essentially unchanged.  The variation shown in Fig.~9b should be correct.)

Our resummed cross sections at $\sqrt{S} = 1.8$ TeV are about $9\%$ above their
next-to-leading order counterparts computed with the same parton
distributions.  To gain numerical insight into the magnitude
of this increase, we
may examine the growth of the cross section in the dominant $q\bar{q}$ channel
that would be expected in a series of fixed-order calculations if only
leading-logartihmic threshold contributions are included in successive orders
in $\alpha$.  We take the universal
leading-logarithmic contributions at ${\cal O}(\alpha)$ above the Born level 
from the full
next-to-leading order calculations of $t\bar{t}$ production and the
leading-logarithmic contributions at ${\cal O}(\alpha^2)$ from
calculations of the Drell-Yan cross section.  For this exercise only, we
set aside our perturbative constraint, Eq.~(\ref{thnineteen}), and instead
we integrate over the entire threshold region $0 < z < 1 $.  We present
the resulting finite order physical cross sections in the $\overline{{\rm MS}}$
scheme in Fig.~10.  (For the Born cross section, we integrate over all $z$.)
We observe that the cross section $\sigma^{(0+1)}$ obtained from the
leading-logarithm terms, only, through ${\cal O}(\alpha)$ is in remarkable
agreement with the cross section obtained from our full next-to-leading order
calculation for $t\bar{t}$ production.  Moreover, our predicted resummed 
cross section lies part way between $\sigma^{(0+1)}$ and the 
cross section $\sigma^{(0+1+2)}$ obtained from the leading-logarithm terms, 
only, through ${\cal O}(\alpha^2)$.  At $m =$ 175 GeV, the increase of 
$\sigma^{(0+1)}$ over the Born result is a 22\% effect, and the further 
increase of $\sigma^{(0+1+2)}$ over $\sigma^{(0+1)}$ is another 14\% effect.  We
conclude that the roughly $9\%$ increase of our final resummed cross section
above the next-to-leading order cross section is quite reasonable.  It would be
surprising if it were much less.

Table 1: The total $t\bar{t}$ production cross section at $\sqrt{S} =$ 1.8 TeV
and its perturbative uncertainty. The theoretical error band, normalized
with respect to the central value, represents an almost constant uncertainty
of 9 to 10$\%$.

\begin{center}
\begin{tabular}{|  c | c | c | c  |}\hline
$m$ (GeV) & $\sigma_{pert}^{t\bar{t}}$ (min; pb) &
$\sigma_{pert}^{t\bar{t}}$ (central; pb) &
$\sigma_{pert}^{t\bar{t}}$ (max; pb) \\ \hline\hline
150 & 11.76 & 12.72 & 12.90 \\ \hline
155 & 9.87 & 10.68 & 10.83 \\ \hline
160 & 8.33 & 9.01 & 9.14 \\ \hline
165 & 7.06 & 7.63 & 7.73 \\ \hline
170 & 6.00 & 6.48 & 6.57 \\ \hline
175 & 5.10 & 5.52 & 5.59 \\ \hline
180 & 4.36 & 4.71 & 4.78 \\ \hline
185 & 3.73 & 4.04 & 4.09 \\ \hline
190 & 3.20 & 3.46 & 3.51 \\ \hline
195 & 2.75 & 2.98 & 3.02 \\ \hline
200 & 2.37 & 2.57 & 2.60 \\ \hline
205 & 2.04 & 2.21 & 2.24 \\ \hline
210 & 1.77 & 1.91 & 1.94 \\ \hline
215 & 1.53 & 1.65 & 1.68 \\ \hline
220 & 1.32 & 1.43 & 1.45 \\ \hline
225 & 1.15 & 1.24 & 1.26 \\ \hline
230 & 0.99 & 1.08 & 1.10 \\ \hline
235 & 0.86 & 0.94 & 0.96 \\ \hline
240 & 0.75 & 0.81 & 0.83 \\ \hline
245 & 0.65 & 0.71 & 0.72 \\ \hline
250 & 0.57 & 0.62 & 0.63 \\ \hline
\end{tabular}
\end{center}

As remarked earlier in the paper, our resummation includes only the universal
logarithmic structures.  It is reasonable to inquire whether and by how much
the predicted cross section would change if subleading logarithmic structures
are included.  One good representation of the possible effect
of subleading structures is the full Drell-Yan
exponent itself, Eq.~(\ref{mutwo}). As shown in Fig.~4, this exponent is
larger than our leading exponent, but the corresponding perturbative regime,
calculated through Eq.~(\ref{thnineteen}), is smaller.
To obtain a cross section that includes the Drell-Yan subleading
logarithms we use Eqs.~(\ref{thone})-(\ref{thtwop}) with
the sum over $Q_j$'s replaced by the full first term, $Q_0$. It is
shown as the dashed line in Fig.~11 for the $q\bar{q}$ channel. We see that
the curves differ little from each other, and that the Drell-Yan resummation
prediction is within our uncertainty band. This example substantiates our
belief that $\mu$-variation is an
adequate measure of perturbative uncertainty and that it includes the effect
of non-resummable, non-universal logarithms.

A second source of uncertainty, that is partly phenomenological, partly
perturbative, and partly correlated with non-universal logarithms, is
associated with the use of different parton distributions.  The parton set we
use is a next-to-leading order determination of the quark and gluon densities.
Since we use resummed partonic cross sections, it is arguably true that we
should also use parton densities based on resummed expressions for deep
inelastic lepton scattering cross sections and other processes used in
the determination of the densities.  However, no such densities exist.  It is
common practice phenomenologically to repeat calculations with different sets of
parton and to estimate thereby a second source of uncertainty.  Except for
the fact that different data sets are used, or are emphasized differently, in
different determinations of parton densities, and that the fitting programs
differ, we opine that the practice of adding in quadrature, or otherwise,
uncertainties associated with $\mu$ variation and those associated with
different parton sets involves significant double counting.  Much of the
difference
among modern parton sets reduces to a difference in $\Lambda$ which largely
affects $\alpha(\mu)$.  Thus, this difference is correlated with the
$\mu$-variation we consider above, and should not be treated as an
independent error.  Over the range $\mu/m\in\{0.5,2\}$, the band
of variation of the strong coupling strength $\alpha_s$ is a
generous $\pm10$\%\ at $m$ = 175 GeV.

In Fig.~9c we present our predictions for an upgraded Tevatron operating
at $\sqrt{S}=2$ TeV. Our cross section is larger than the next-to-leading
order one by about $9\%$.  We predict
\begin{equation}
\sigma^{t\bar{t}}(m=175\ {\rm GeV},\sqrt{S}=2\ {\rm TeV})=
7.56^{+0.10}_{-0.55}\ pb\ .
\label{predthree}
\end{equation}
At $m = 175$ GeV, the value of the cross section at $\sqrt{S} = 2$ TeV is
about 37\%\ greater than that at $\sqrt{S}= 1.8$ TeV.

Turning to $pp$ scattering at the energies of the Large Hadron Collider (LHC)
at CERN, we note a few significant differences from $p\bar{p}$ scattering at the
energy of the Fermilab Tevatron.  The dominance of the $q {\bar q}$ production
channel at the Tevatron is replaced by $g g$ dominance at the LHC.  Owing to
the much larger value of $\sqrt{S}$, the near-threshold region in the subenergy
variable is relatively less important, reducing the significance of
initial-state soft gluon radiation.  Lastly, physics in the region of large
partonic subenergy $\sqrt{s}$, where straightforward next-to-leading order
QCD is also inadequate\cite{ref:collins}, may become more significant for
$t\bar{t}$ production at LHC energies than the effects of initial-state
radiation.  The approach of our paper is limited to the resummation of
initial-state gluon radiation only.  We present estimates in Fig.~9d of
the cross section for LHC energies of 10 and 14 TeV. We obtain
\begin{equation}
\sigma^{t\bar{t}}(m=175\ {\rm GeV},\sqrt{S}=14\ {\rm TeV})=760\ pb\ .
\label{predfour}
\end{equation}
\subsection{Comparisons with other calculations}

In earlier sections we comment on differences between our formalism and that
of Laenen, Smith, and vanNeerven (LSvN) \cite{ref:laeneno, ref:laenent}.  Here
we compare aspects of our numerical predictions.  The comparison most
relevant to experiment is that of our Table~1 and the corresponding table in
Ref.~\cite{ref:laenent}.  Our central values are 10 to 14\% larger (the
difference increases with mass), and our estimated theoretical uncertainty is
9 to 10\% compared with their 28\% to 20\% (decreasing with mass).  The two
predictions have overlapping uncertainties and are, in this sense, in
agreement.  In commenting on differences, we remark that our
Born cross section is about 3 to 5\% larger than LSvN's Born cross section.
The difference arises from the different parton distributions used in the
two calculations, including differences in $\Lambda$ which alone account
for half or more of this increase.  This source of difference should be kept
in mind when comparisons are made with next-to-leading order calculations and
various previous results.

It is important to stress that the theoretical uncertainties are
estimated in quite different ways in the two methods.
We use the standard $\mu$-variation, whereas LSvN obtain their uncertainty
primarily from variations of their undetermined IR cutoffs.
From a theoretical point of view, a study of the variation of the cross
section with the hard-scale $\mu$ is important because it deals with
stability of the calculation under variation of a perturbative, but not directly
determinable renormalization/factorization scale. This statement applies as
well to the LSvN calculation, above and beyond the choice of their IR
cutoff.  The role of the IR cutoff is to measure ignorance of non-perturbative
effects in the LSvN approach.  One of the advantages
of a resummation calculation should be diminished dependence
of the cross section on $\mu$, less variation than is present in fixed-order
calculations.

It is also instructive to compare the numerical values of our
perturbative boundary, $\eta_0(m,\mu/m)$, with the corresponding cut in $\eta$
produced by LSvN's IR cutoff $\mu_0$.  In the $\overline{{\rm MS}}$ scheme,
their cut is\cite{ref:laeneno}
\begin{equation}
\eta(\mu_0,\mu)={1\over 2}\biggl({\mu_0\over \mu}\biggr)^3 ,
\label{irrelevantsix}
\end{equation}
and in the DIS scheme, it is
\begin{equation}
\eta(\mu_0,\mu)={1\over 2}\biggl({\mu_0\over \mu}\biggr)^2 .
\label{irrelevantsev}
\end{equation}
Values of $\eta(\mu_0,\mu)$ are provided in Tables~2 and 3 and
are compared with
our perturbative boundary $\eta_0(m,\mu/m)$.

In the paragraphs to follow, we compare $\mu$-variation, scheme
dependence, and the influence of the difference between our perturbative
boundary and LSvN's cut in $\eta$.  In Tables~2 and 3, we reproduce numbers
from Ref.~\cite{ref:laeneno} and show the corresponding values we calculate.
The comparison is made at $m= $100 GeV in Table~2 because it is at this value
of the top mass that they provide results in the $\overline{{\rm MS}}$ scheme
for both the $q\bar{q}$ and $g g$ channels.

\newpage
Table 2: Physical cross sections in pb for $m=100$ GeV at $\sqrt{S} = $
1.8 TeV.  The LSvN predictions are shown for various choices of their IR cutoff
$\mu_0$.
Our perturbative boundary $\eta_0$ and LSvN's phase space cut $\eta(\mu_0,\mu)$
are also shown. Absences in the LSvN entries denote extremely large numbers.

\begin{center}
\begin{tabular}{|  c | c | c | c  |}\hline
$\sigma_{q\bar{q}}$($m=100$ GeV; $\overline{{\rm MS}}$)
& $\mu/m=0.5$ & $\mu/m=1$ & $\mu/m=2$ \\ \hline\hline
Born & 64.3 & 48.4 & 37.3 \\ \hline
NLO & 54.6 & 55.0 & 51.3 \\ \hline
$\sigma_{q\bar{q}}^{pert}$ & 56.9 & 59.4 & 57.2 \\ \hline
LSvN($\mu_0=0.1$m) & 52.6 & 88.9 & - \\ \hline
LSvN($\mu_0=0.2$m) & 42.5 & 66.4 & 110.5 \\ \hline
$\eta_0(\mu)$ & 0.007 & 0.011 & 0.014 \\ \hline
$\eta(\mu_0=0.1{\rm m},\mu)$ & 0.004 & 0.0005 & 0.000063 \\ \hline
$\eta(\mu_0=0.2{\rm m},\mu)$ & 0.032 & 0.004 & 0.0005 \\ \hline\hline
$\sigma_{gg}$($m=100$ GeV; $\overline{{\rm MS}}$)
& $\mu/m=0.5$ & $\mu/m=1$ & $\mu/m=2$ \\ \hline\hline
Born & 36.4 & 23.8 & 16.1 \\ \hline
NLO & 44.5 & 40.5 & 34.0 \\ \hline
$\sigma_{gg}^{pert}$ & 44.8 & 41.6 & 35.0 \\ \hline
LSvN($\mu_0=0.2$m) & 34.3 & 88.7 & - \\ \hline
LSvN($\mu_0=0.3$m) & 28.1 & 42.4 & 820.0 \\ \hline
$\eta_0(\mu)$ & 0.036 & 0.063 & 0.1 \\ \hline
$\eta(\mu_0=0.2{\rm m},\mu)$ & 0.032 & 0.004 & 0.0005 \\ \hline
$\eta(\mu_0=0.3{\rm m},\mu)$ & 0.108 & 0.0135 & 0.0017 \\ \hline\hline
$\sigma_{t\bar{t}}$($m=100$ GeV; $\overline{{\rm MS}}$)
& $\mu/m=0.5$ & $\mu/m=1$ & $\mu/m=2$ \\ \hline\hline
Born & 100.7 & 72.2 & 53.4 \\ \hline
NLO & 99.1 & 95.5 & 87.4 \\ \hline
$\sigma_{t\bar{t}}^{pert}$ & 101.7  & 101.0 & 92.2 \\ \hline
LSvN($\mu_0=\{0.2,0.3\}$m) & 70.6 & 108.8 & 930.5 \\ \hline
\end{tabular}
\end{center}

\newpage
Table 3: Physical cross sections in pb and perturbative boundaries at
$m=150$ GeV and $\sqrt{S} = 1.8$ TeV.  The corresponding LSvN predictions are
also shown.

\begin{center}
\begin{tabular}{|  c | c | c | c  |}\hline
$\sigma_{q\bar{q}}$($m=150$ GeV; ${\rm DIS}$) & $\mu/m=0.5$ & $\mu/m=1$ & $\mu/m=2$ \\ \hline\hline
NLO & 9.42 & 9.31 & 8.57 \\ \hline
$\sigma_{q\bar{q}}^{pert}$ & 9.76 & 9.92 & 9.31 \\ \hline
LSvN($\mu_0=0.1$m) & 7.9 & 10.0 & 9.7 \\ \hline
$\eta_0(\mu)$ & 0.0017 & 0.0024 & 0.0028 \\ \hline
$\eta(\mu_0=0.1{\rm m},\mu)$ & 0.02 & 0.005 & 0.00125 \\ \hline\hline
$\sigma_{gg}$($m=150$ GeV; $\overline{{\rm MS}}$) &
$\mu/m=0.5$ & $\mu/m=1$ & $\mu/m=2$ \\ \hline\hline
NLO & 2.51 & 2.22 & 1.81 \\ \hline
$\sigma_{gg}^{pert}$ & 2.53 & 2.30 & 1.89 \\ \hline
LSvN($\mu_0=0.2$m) & 1.76 & 4.38 & - \\ \hline
$\eta_0(\mu)$ & 0.033 & 0.055 & 0.083 \\ \hline
$\eta(\mu_0=0.2{\rm m},\mu)$ & 0.032 & 0.004 & 0.0005 \\ \hline
\end{tabular}
\end{center}

Tables~2 and 3 show that our resummed cross sections satisfy the test of
stability under variation of the hard scale $\mu$.  The resummed results show
less variation than the next-to-leading order cross section. On the other
hand, this is not true of the resummation of Ref.~\cite{ref:laeneno}.
This distinction is linked to the absence of undetermined
IR cutoffs in our method and the specific RG-invariant exponent we use.
Both of these differences contribute to the instability apparent in the results
of Ref.~\cite{ref:laeneno}.  The LSvN results show less variation with $\mu$
in the DIS scheme than in the $\overline{{\rm MS}}$ scheme, the reason they
provide their final predictions in the DIS scheme.  As shown in Table~3, the
variation with $\mu$ of their $q\bar{q}$ cross section in the DIS scheme is
about $21\%$. For comparison, the next-to-leading order cross section shows a
variation of $9\%$ and our resummed cross section a variation of $6\%$.

Our perturbative boundary, $\eta_0(\mu)$, is fairly insensitive
to $\mu$-variation.  In the $\overline{{\rm MS}}$ scheme, it changes by
about $30\%$ around its central value (for $\mu=m$) while $\mu$ itself
changes by 50 to 100\% from the central value.
In the DIS scheme the changes in $\eta_0(\mu)$ are even smaller, ranging from
16 to 28\%. The mild changes in $\eta_0(\mu)$ accord with our earlier physical
expectation that the boundary is characteristic of the particle's mass,
not of the perturbative artifact $\mu$.
If a complete resummation were possible, as in the Drell-Yan
case shown in Fig.~6, our boundary $\eta_0$ would be a function
of mass only.  We may contrast the modest changes of our $\eta_0(\mu)$ with the
fact that the IR boundary of LSvN varies by three orders of magnitude
in the $\overline{{\rm MS}}$ scheme.  This large change is partly responsible
for the unstable behavior of LSvN's cross section with $\mu$.
In the DIS scheme, the IR boundary varies
by an order of magnitude for $\mu_0=0.1m$ (the value used for their
central value prediction in Ref.~\cite{ref:laenent}), and at $\mu/m=1$ it
is twice as large as our $\eta_0$.
The fact that these quantities are of the same order of magnitude
makes LSvN's and our final predictions for the $q\bar{q}$ cross section
comparable. Since the $q\bar{q}$ channel is dominant, our final predictions
for the total $t\bar{t}$ cross section at the Tevatron are also equal within
uncertainties.

Scheme dependence is an extra source of theoretical uncertainty, but it
should produce minimal differences for physical cross sections.  The results 
in Tables~2 and 3 show that this is not true of the the LSvN cross sections.  
We provide our main predictions in the $\overline{{\rm MS}}$
factorization scheme\cite{ref:edpapero}.  To check for possible scheme
dependent uncertainty, we perform our resummation for the dominant
$q\bar{q}$ channel in both schemes.  The cross sections presented in Table~4
show that scheme dependence is insignificant in our approach, resulting in
a difference of about $4\%$ for the cross section.

Table 4. Physical cross sections in pb for the $q\bar{q}$ channel: DIS versus
$\overline{{\rm MS}}$ scheme.

\begin{center}
\begin{tabular}{|  c | c | c | c | c  |}\hline
$m$ (GeV)& $\sigma_{q\bar{q}}$({\rm DIS}; $\overline{{\rm MS}}$) & $\mu/m=0.5$ & $\mu/m=1$ & $\mu/m=2$ \\ \hline\hline
100 & NLO & 52.3; 54.6 & 52.4; 55.0 & 48.9; 51.3 \\ \hline
    & $\sigma_{q\bar{q}}^{pert}$ & 53.9; 56.8 & 55.5; 59.4 & 52.8; 57.2 \\ \hline
125 & NLO & 21.1; 21.9 & 21.0; 21.8 & 19.5; 20.1 \\ \hline
    & $\sigma_{q\bar{q}}^{pert}$ & 21.9; 22.9 & 22.3; 23.7 & 21.1; 22.6 \\ \hline
150 & NLO & 9.42; 9.68 & 9.31; 9.53 & 8.57; 8.73 \\ \hline
    & $\sigma_{q\bar{q}}^{pert}$ & 9.76; 10.16 & 9.92; 10.42 & 9.31; 9.87 \\ \hline
175 & NLO & 4.46; 4.54 & 4.39; 4.43 & 4.01; 4.02 \\ \hline
    & $\sigma_{q\bar{q}}^{pert}$ & 4.63; 4.78 & 4.69; 4.87 & 4.37; 4.58 \\ \hline
200 & NLO & 2.20; 2.21 & 2.15; 2.14 & 1.96; 1.93 \\ \hline
    & $\sigma_{q\bar{q}}^{pert}$ & 2.29; 2.34 & 2.30; 2.37 & 2.14; 2.21 \\ \hline
\end{tabular}
\end{center}

In a very recent paper\cite{ref:jerks} doubts are expressed about the numerical
importance of resummation for top quark production at Fermilab Tevatron
energies, and criticisms are leveled at our formalism and that of
Ref.~\cite{ref:laeneno}.  We consider the criticisms unfounded.  The 
analysis presented in our current paper substantiates the work we
presented in Ref.~\cite{ref:edpapero}.  As demonstrated in section 3,
our perturbative resummation exponent, e.g., Eq.~(\ref{tseventeen}),
contains no factorially growing terms in
its expansion.  The analysis we present of our Figs.~1 through 5 shows
that the perturbative region in which we apply resummation remains
far removed from the part of phase space in which renormalon poles or
non-perturbative residual uncertainty could be influential.
Upon expansion in terms of the QCD coupling strength,
fixed at the scale of the top quark mass, our formalism produces the universal
leading logarithmic structures that are found at next-to-leading
order (one-loop) in top quark production and at two-loops in the
Drell-Yan process.  This is an all-orders expansion, but it is convergent
because its coefficients do not have factorial growth, and the momentum scale
of these structures is restricted to our calculable perturbative regime. 
Our approach is consistent with, and makes maximum use of, the
perturbative properties of QCD and established finite-order calculations
for the relevant processes.  With respect to the numerical importance of
resummation, we remark that the enhancement of the cross section produced by
resummation in our calculation is a modest 10\% above the next-to-leading
order result, at a top mass of 175 GeV and $\sqrt {S} = 1.8$ TeV.  That the
increase cannot be much less than 10\% is suggested simply by an examination of
the leading-logarithmic contributions at the two-loop level, described in
Fig.~10.
As shown first by LSvN\cite{ref:laeneno}, it is not necessary to transform to
moment space or to resum in order to obtain an increase of the cross section in
the neighborhood of 10\%.

\section{Estimates of Non-Perturbative contributions}

In this section we hazard some estimates of the non-perturbative
contributions. This discussion is based on educated guesses
founded mainly on our intuition of what a physical threshold is
as well as on the behavior of parton distributions near the threshold.
Since even the gluon parton distributions are not altogether well established,
our estimates may be viewed as rough physical understanding expressed
quantitatively.

An examination of Fig.~8 indicates that the non-perturbative regime is
very small, especially for the $q\bar{q}$ channel.  We focus on this channel
first, where the support of the non-perturbative interval is
$\eta_0\simeq 0.008$.  We recall that the area under the next-to-leading order
curve is already included in our resummed cross section.  Concentrating
on the solid curve, we assume that the threshold behavior in the
non-perturbative regime is a finite continuation of the solid curve to
$\eta =0$.  Reasonable guesses include a continuation of the
solid curve as a constant from its peak value down to $\eta=0$,
or, at most, as a smooth function having the same slope as the solid
curve just above the peak.  In the former scenario we get an extra
contribution
\begin{equation}
\delta\sigma_{q\bar{q}}\simeq 5\times 0.008=0.04\ pb.
\label{irrelevanteigh}
\end{equation}
In the latter case we add a further
\begin{equation}
\delta '\sigma_{q\bar{q}}\simeq {1\over 2}(15-10)\times 0.008=0.02\ pb.
\label{irrelevantnin}
\end{equation}
The total is
\begin{equation}
\Delta\sigma_{q\bar{q}}\simeq 0.06\ pb.
\label{irrelevanttwen}
\end{equation}
Since the perturbative cross section at $m = 175$ GeV is
$\sigma_{q\bar{q}}^{pert}=4.87$, the extra contribution amounts to an
increase of about 1 to 1.5$\%$.

For the $gg$-channel displayed in Fig.~8b, $\eta_0\simeq 0.05$, and we find
\begin{equation}
\Delta\sigma_{gg}\simeq {1\over 2}(2-1)\times 0.05=0.025\ pb.
\label{irrelevanttwoeon}
\end{equation}
Given that $\sigma_{gg}^{pert}\simeq 0.65$ pb, this correction is a $4\%$
effect. We estimate that the total non-perturbative correction
cannot be more than
\begin{equation}
\Delta\sigma^{t\bar{t}}\simeq 0.02\sigma^{t\bar{t}}_{pert}.
\label{irrelevanttwetw}
\end{equation}
Correspondingly, we can expect a maximum cross section of about $5.7$ pb
at $m=175$ GeV.

\section{Discussion and conclusions}

There are two main aspects of this paper, one of a more theoretical nature
and the other phenomenological.  First, we describe in all generality
features of {\it perturbative} resummation that occur in a variety of hard
scattering processes while focusing on the specific
process of $t\bar{t}$ production. This process has the advantage of
being physically very interesting in pQCD in that it probes our
understanding of the theory when {\it multiple}
physical scales are involved. It is characterized by an unquestionably large
physical scale, $m$,  and a {\it partonic} variable $\eta$ that, in turn, is the
ratio of two large scales.  It is an ideal process in which one may
examine the systematics of partonic interactions
in more detail than previously possible. In this process, we
encounter difficulties that have challenged the applicability of pQCD for a
long time, notably IR renormalons and large color factors, as well as
the systematics of universality and invariance under changes of the hard scale.
In dealing with $t\bar{t}$ production, we treat all these issues in
a {\it concrete}, phenomenologically significant way.

Second, the paper is concerned with predictions for the total $t\bar{t}$
cross section and the practical application of the theoretical resolutions
developed in the first part. At the same time, comparisons are made with
next-to-leading order cross sections and with earlier resummation
calculations, and extensive discussions are presented of theoretical
limitations and uncertainties.

Our theoretical analysis shows that perturbative resummation without
a model for non-perturbative behavior is both {\it possible} and
advantageous.  In perturbative resummation, the perturbative region of phase
space is separated cleanly from the region of non-perturbative behavior.
The former is the region where large threshold corrections exponentiate
but behave in a way that is {\it perturbatively stable}.
The asymptotic behavior of the QCD perturbative series, including
large multiplicative color factors, is flat, and excursions
around the optimum number of perturbative terms does not create
numerical instabilities or intolerable RG-dependence.  Infrared
renormalons are far away from the stability plateau and, even though
their presence is essential for defining this plateau, they are
of no numerical consequence in the perturbative regime.
Large color factors, which are multiplicative, enhance the IR renormalon
effects and contribute significantly to limiting the perturbative regime.

As expected in a process with two physical scales,
the constraint that insures good perturbative behavior is a function of
several parameters. Making the customary assumption that
the renormalization scale is identified with the factorization scale
(we denote this single scale by $\mu$), and denoting the variable
that probes the partonic threshold by $\eta$, we may summarize the essence of
our perturbative resummation procedure as follows:

The threshold logarithmic corrections are exponentiated in an exponent
\begin{equation}
E=E(x,\alpha(\mu),m/\mu,N(\mu))
\end {equation}
where $x=\ln(1/(1-z))$.  If the total center-of-mass energy $\sqrt{S}$ is not
far above threshold,  $\eta\simeq S/4m^2 -1\simeq (1-z)/2$.
The number of perturbative terms $N(\mu)$ depends on the hard scale only,
as long as IR renormalons are far away.  The
cross section is insensitive to fine-tuning of this number, as long as
we remain in a specific well-defined perturbative region of $x$.
In this paper, we determine this region in momentum space by demanding that the
cross section conform to {\it perturbative} power-counting.  The cross
section is a series multiplying the exponentiated effects
of the threshold and whose successive terms are given by functions of
the successive derivatives $\partial^k E/k!\partial x^k$.  Perturbative
power-counting means that higher derivatives
are suppressed relative to lower ones, because they contribute
subleading logarithmic structures in $\ln(1/(1-z))$. This statement
of perturbative power-counting
is the essence of perturbation theory.  Imposing its validity
in the two-scale parameter space, we determine the perturbative regime:
\begin{equation}
{\partial\over \partial x}E(x,\alpha(\mu),m/\mu,N(\mu))\le 1\ .
\label{irrelevanttweth}
\end{equation}
This constraint restricts several sources of non-perturbative
behavior, namely: IR renormalons, $\alpha(\mu)$, $x$, and color factors.
Some of them are channel-independent while others depend on the
nature of the interacting partons. The constraint serves to identify
the only unknown in the problem, i.e., the momentum range in $x$ in which
perturbative resummation is justified.

We calculate
\begin{equation}
\sigma^{t\bar{t}}(m=175\ {\rm GeV},\sqrt{s}=1.8\ {\rm TeV})=
5.52^{+0.07}_{-0.42}\ pb\ .
\label{predtwo}
\end{equation}
Our total $t\bar{t}$-production cross section at $m =$ 175~GeV and
$\sqrt{S} = 1.8$ TeV is 10 to 14\% greater than that of earlier calculations.
Part of the increase comes from the more recent parton distributions we use.
Our resummed cross sections are about $9\%$ above the next-to-leading order
cross sections computed with the same parton distributions. The
renormalization/factorization scale dependence of our cross section
is fairly flat, resulting in a 9 to 10\% theoretical uncertainty.  This
variation is smaller than the corresponding dependence of the next-to-leading
order cross section, and it is much smaller than the corresponding dependence
of the resummed cross section of Ref.~\cite{ref:laeneno}.
There are other perturbative uncertainties, such as dependence on
parton distributions and factorization scheme. Each of these sources
affects our cross section minimally, at level of $4\%$ or less.
These variations are not altogether independent, so
we opt not to add them in estimating the theoretical uncertainty.
For example, different parton distributions are associated with different
values of $\Lambda$ and therefore of $\alpha_s$. This uncertainty is
correlated with uncertainty in $\alpha_s$ from other measurements, {\it and}
with the standard $\mu$-variation that we use.  Additional back-of-the-envelope
modelling of non-perturbative behavior does not increase our cross section
more than $2\%$. However, we do not claim to address the issue of
possible non-perturbative enhancements other than at the level of a
conservative educated guess.

Our theoretical analysis and the stability of our cross sections under $\mu$
variation provide confidence that our perturbative resummation procedure
yields an
accurate calculation of the inclusive top quark cross section at Tevatron
energies and exhausts present understanding of the perturbative content
of the theory. Our prediction agrees with data, within the large
experimental uncertainties. We anticipate the greater precision of top quark
production data from future Tevatron collider runs will be instrumental in a
fundamental test of perturbative QCD.  The data
may also guide the modelling of the non-perturbative phase of the theory and
provide a glimpse of phenomena beyond the Standard Model.

The methodology described in this paper can be applied in several closely
related situations.  The production of bottom quarks in hadron
reactions at energies typical of the Fermilab fixed-target program or the
HERA-B facility should be sensitive to the same type of threshold
enhancements.  Other reactions include the production of
hadronic jets with very large values of transverse momentum or the production
of very massive lepton pairs in Tevatron collider experiments.  We hope to
address these topics in the near future.

\centerline{\bf Acknowledgments}

We thank Dr. Sissy Kyriazidou and Dr. Stephen Mrenna for helpful discussions.
This work was supported in part by the U.S. Department of
Energy, Division of High Energy Physics, contract No. W-31-109-ENG-38.



\appendix{\bf FIGURE CAPTIONS}

\begin{description}
\item{Figure 1.} Optimum number of perturbative terms in the exponent.
\begin{description}
\item{(a)} Normalized principal-value exponents (solid) and their perturbative
approximations (dashed) as a function of $N$, for fixed mass and for four
parametric moment values (from the bottom, at $n=10,20,30,40$).
Optimization occurs at $N(t(m=175{\rm Gev})) =6$ for all four moments.
\item{(b)} The function $N(t(m))$ (solid) and its simple analytic
approximation $[t(m)-3/2]$ (dashed).
\end{description}
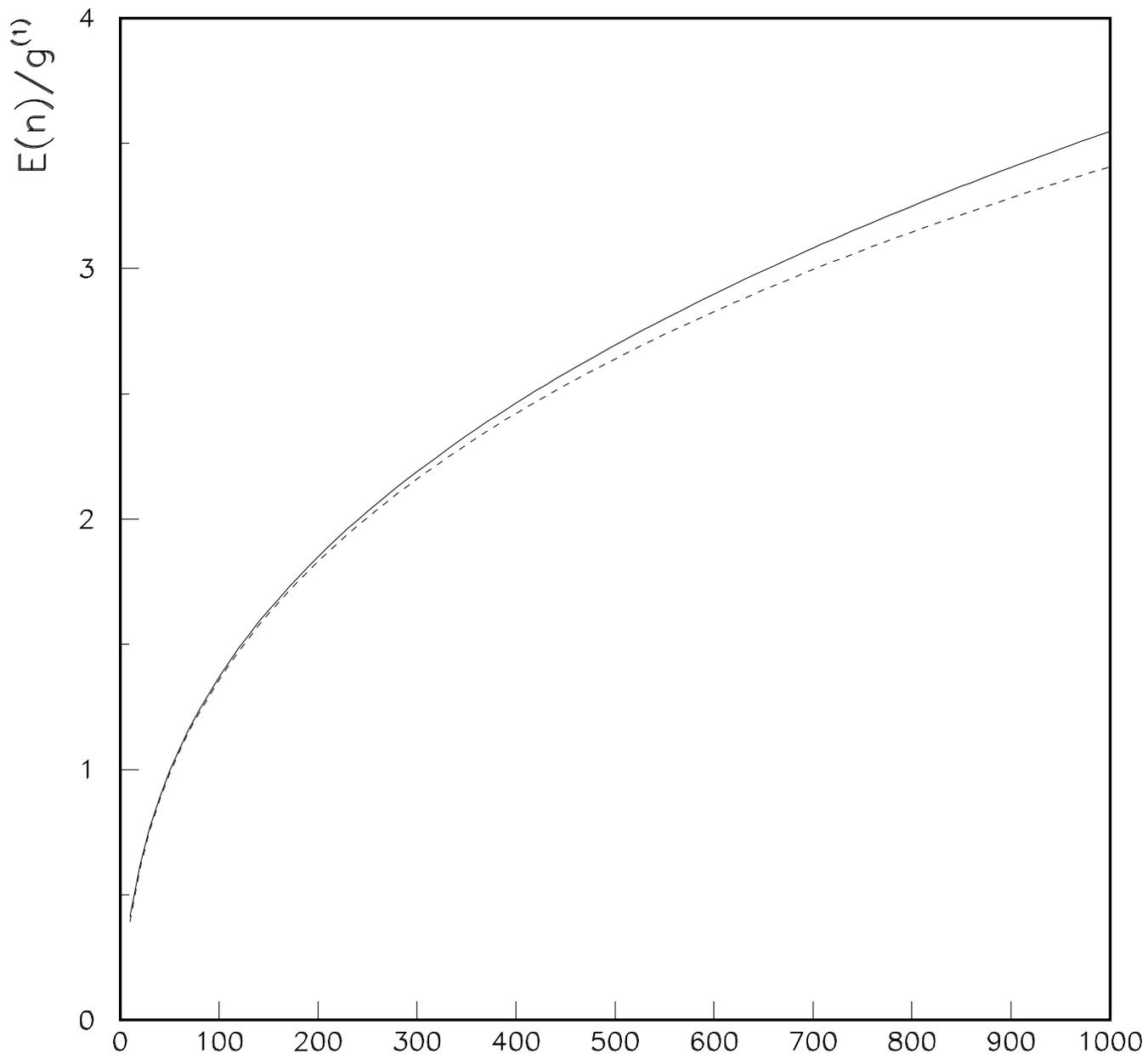
\item{Figure 2.} The normalized principal-value exponent (solid) and its
perturbative approximation (dashed) versus the moment, $n$.
\item{Figure 3.} The normalized perturbative exponent without
regulators:
\begin{description}
\item{(a)} Partial sums versus the number $N$ of
terms as a parametric family in $n$: Solid bunch is for $n\in\{100,1000\}$
(increasing) in steps of $100$. Dotted bunch is for $n=2000,\ 3000$
(increasing). Dashed bunch is for $n=10000,\ 20000,\ 30000$ (increasing).
\item{(b)} Slope function versus $N$ for
$n=10,\ 50,\ 100,\ 500,\ 1000,\ 2000$ (increasing).  The
minimum is attained at $N-1\simeq N(t(m=175{\rm GeV}))=6$, for $n\le 1000\simeq
m/\Lambda $.
\end{description}
\item{Figure 4.} Renormalization/factorization hard scale dependence of the
resummation exponents:
\begin{description}
\item{(a)} Drell-Yan perturbative (solid), principal value (dotted),
and leading-logarithm (dashed)
exponents versus $\mu$ for fixed mass and parametric moment
values $n=10,\ 50,\ 100$ (increasing); $N = N(t(m=175{\rm GeV}))+1$.
\item{(b)} Parametric families of partial sums $N=1,...N(t(m=175{\rm GeV}))+1$
(increasing) versus $\mu$ for the Drell-Yan (solid) and leading logarithm
(dashed) exponents, at $n=50$.
\end{description}
\item{Figure 5.} Evaluation of the perturbative regime:
\begin{description}
\item{(a)} Saturation of the perturbative constraint in the $q\bar{q}$ channel
as derived in
moment space ($S(n)=$ solid, $\alpha_s{\rm e}^{-E(n)}=$ dashed (Drell-Yan),
dotted (leading logarithms)) and in momentum space ($P_1(n)\le 1$, dashed =
Drell-Yan, dotted = leading logarithms); $N = N(t(m=175{\rm GeV}))+1$.  The
upper curves are those in momentum space.
\item{(b)} Same as in (a) but for the $gg$ channel.
\end{description}
\item{Figure 6.} Hard-scale dependence of the perturbative regime:
\begin{description}
\item{(a)} The function $P_1(n)$ vs. $n$ as a parametric family for
$\mu=100,150,200,250,300$ GeV for the Drell-Yan exponent (solid bunch)
and the leading-logarithm exponent (dashed bunch, decreasing)
for the $q\bar{q}$ channel.
\item{(b)} Same as in (a) for the $gg$ channel.
\end{description}
\item{Figure 7.} The partonic cross sections $\sigma(\eta)$ as a function of
$\eta$:
\begin{description}
\item{(a)} Born (dotted), next-to-leading order (dashed), and resummed (solid)
partonic cross sections for the $q\bar{q}$ channel.
\item{(b)} Same as in (a) for the $gg$ channel.
\end{description}
\item{Figure 8.} Differential cross section $d\sigma/d\eta$  as a function of
$\eta$ in the $\overline{{\rm MS}}$ factorization scheme.
The physical cross sections are the areas under the curves.
\begin{description}
\item{(a)} $q\bar{q}$ channel: Born (dotted), next-to-leading order (dashed),
and resummed (solid).
The non-perturbative regime is the area from $\eta =0$ to the point in
$\eta$ at which the solid and dashed curves intersect.
\item{(b)} Same as in (a) for the $gg$ channel.
\end{description}
\item{Figure 9.} Inclusive cross section for top quark production in the
$\overline{{\rm MS}}$ scheme:
\begin{description}
\item{(a)} At the Tevatron, for $p\bar{p} \rightarrow t\bar{t} X$ at
$\sqrt{S}=1.8$ TeV. The extremum dashed lines are our band of
uncertainty, and the solid line between them is our central-value
prediction.
We also reproduce the published data of the CDF and D0 collaborations.
\item{(b)} Hard-scale dependence of the resummed (solid) and next-to-leading
order (dashed) cross sections at $\sqrt{S}=1.8$ TeV for $m=175$ GeV.
\item{(c)} Same as in (a), but for the upgraded Tevatron, $\sqrt{S}=2.0$ TeV.
\item{(d)} Central values of the resummed cross section (solid) for
$p p \rightarrow t\bar{t} X$ at the CERN LHC energies of 10 and 14 TeV and
the corresponding next-to-leading order predictions (dashed).
\end{description}
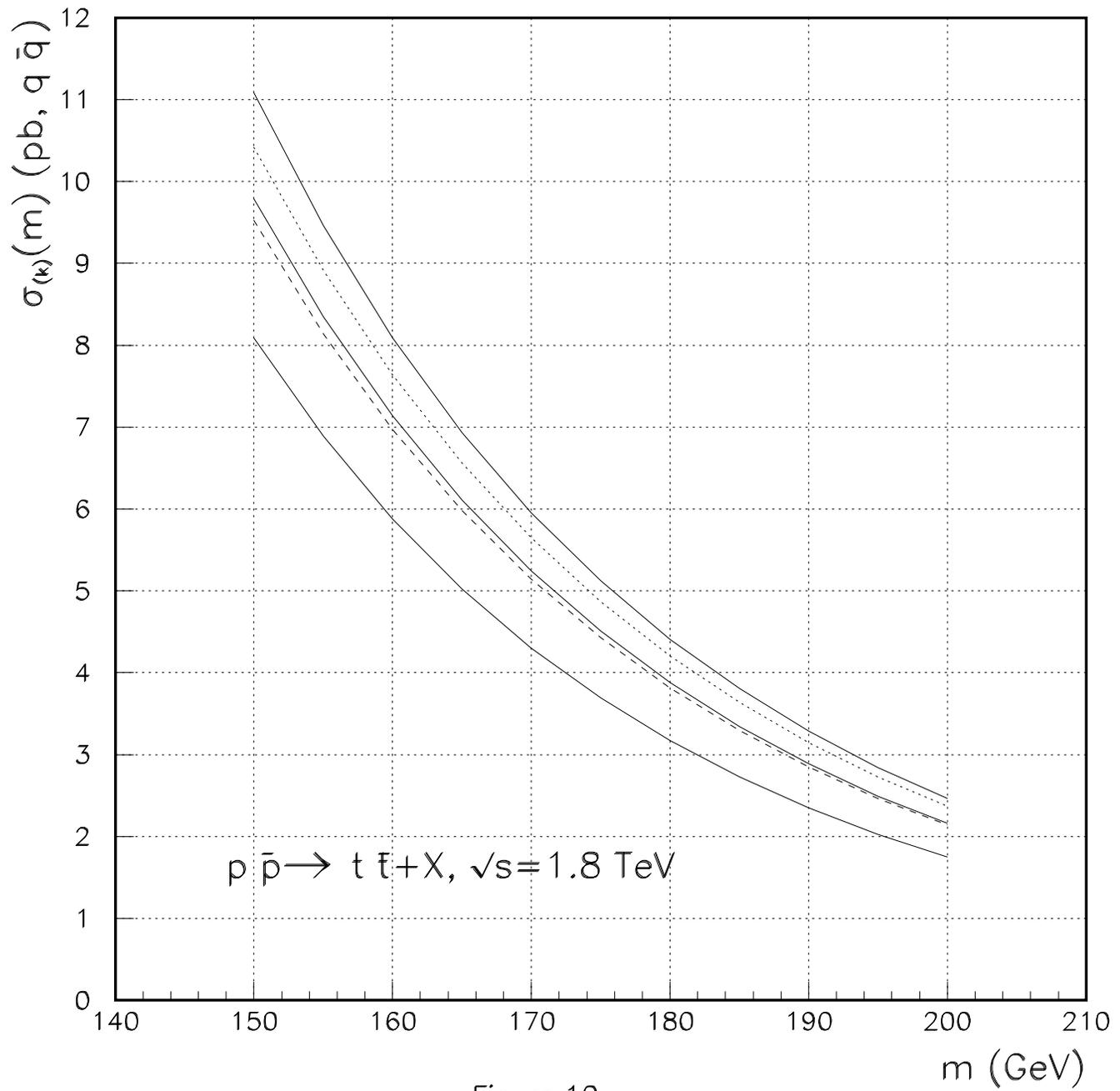
\item{Figure 10.} Physical cross sections in the $q\bar{q}$ channel in
the $\overline{{\rm MS}}$ scheme.
The solid lines denote the finite-order partial sums of the
universal leading-logarithmic contributions from the explicit
${\cal O}(\alpha)$ and ${\cal O}(\alpha^2)$ calculations for the
$t\bar{t}$ and Drell-Yan cross sections, respectively, integrated throughout
phase space.
Lower solid: $\sigma^{(0)}$;
middle solid: $\sigma^{(0+1)}$;
upper solid: $\sigma^{(0+1+2)}$. The dashed curve represents the exact
next-to-leading order calculation for $t\bar{t}$ production,
in excellent agreement with $\sigma^{(0+1)}$. The dotted curve is our
resummed prediction.
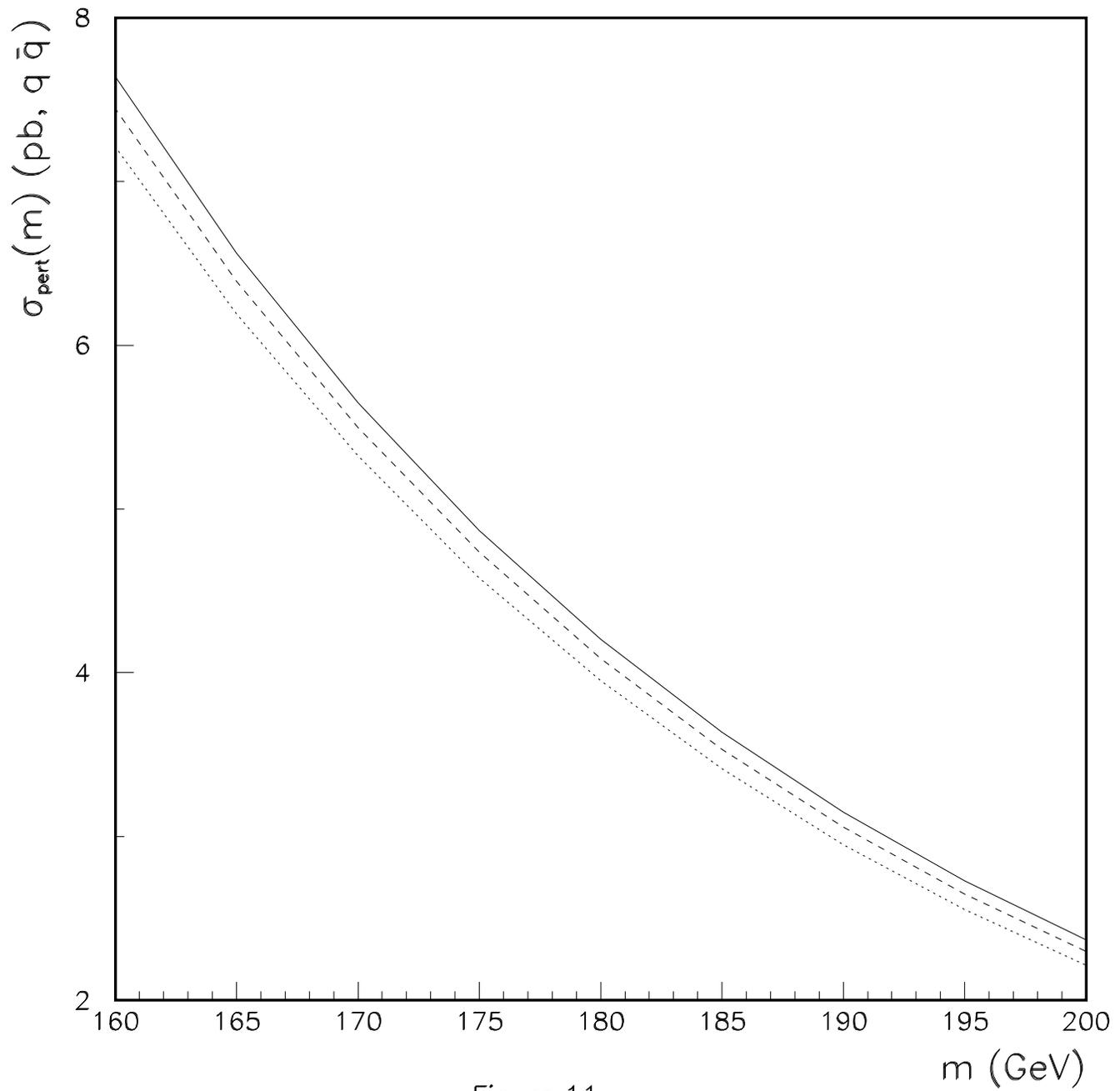
\item{Figure 11.} Physical cross sections for the $q\bar{q}$ channel in the
$\overline{{\rm MS}}$ scheme:
Leading-logarithm resummed cross sections for $\mu/m=1$ (solid) and $\mu/m=2$
(dotted) and the Drell-Yan resummed version for $\mu/m=1$ (dashed).
\end{description}
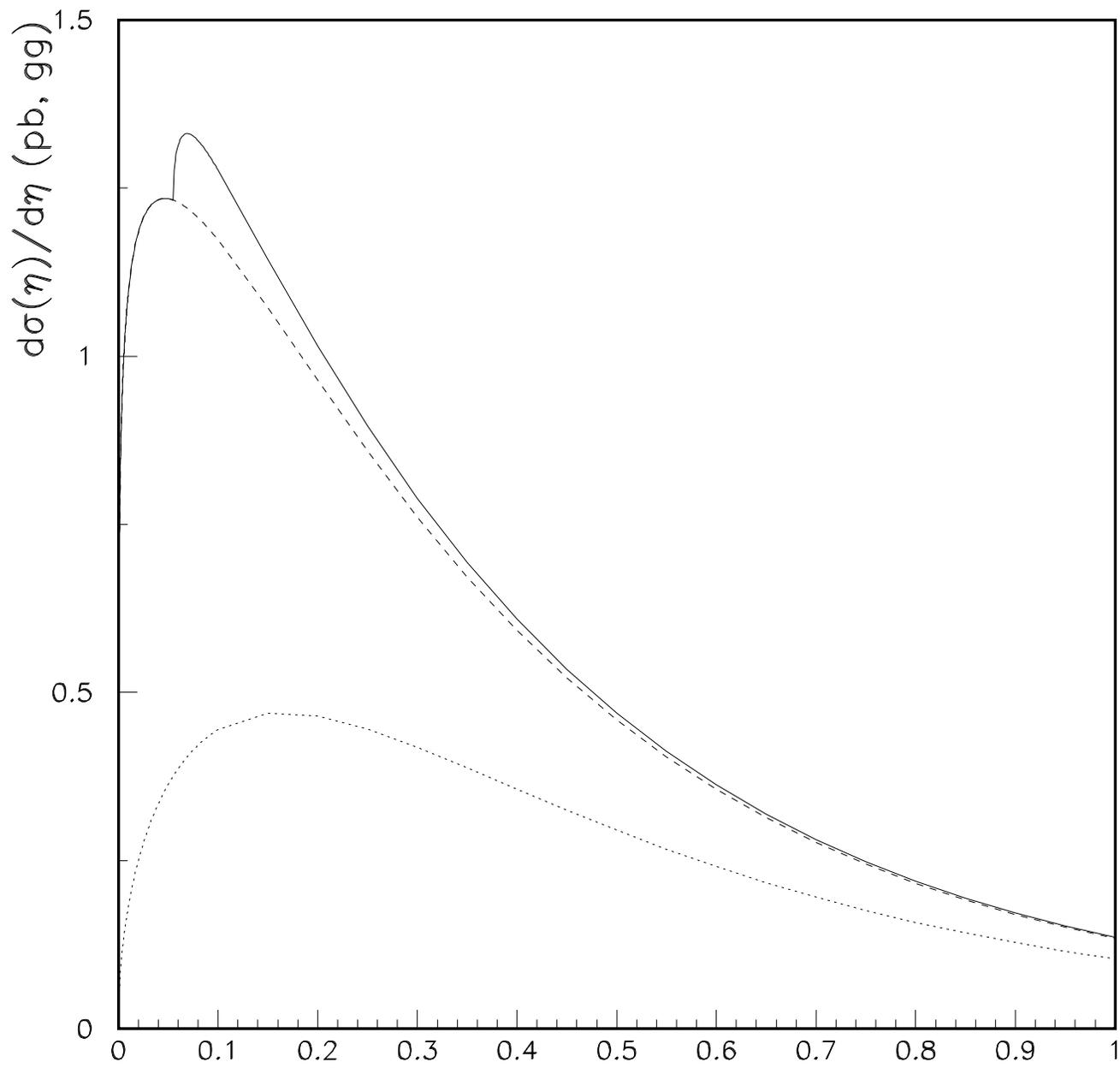
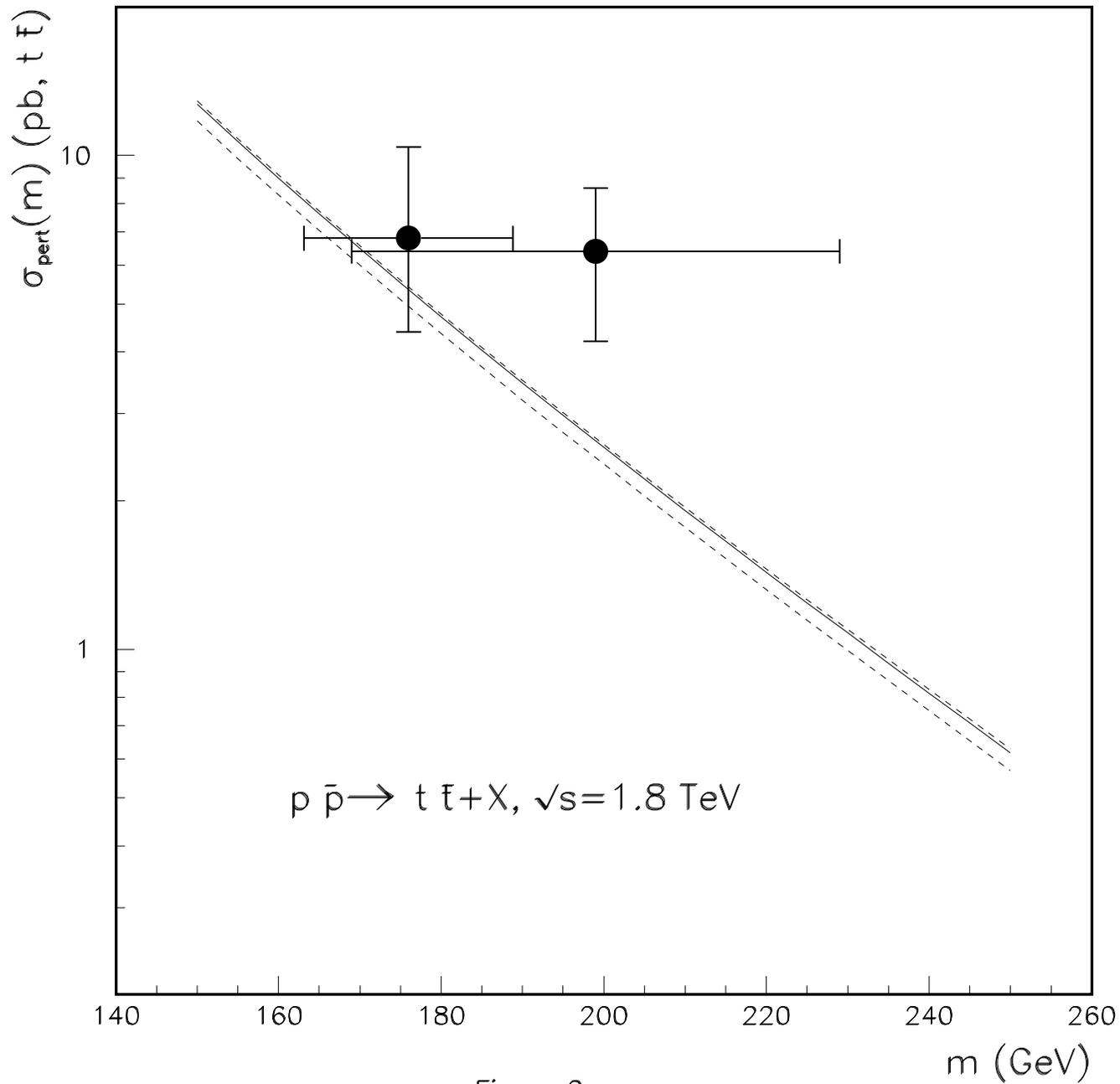
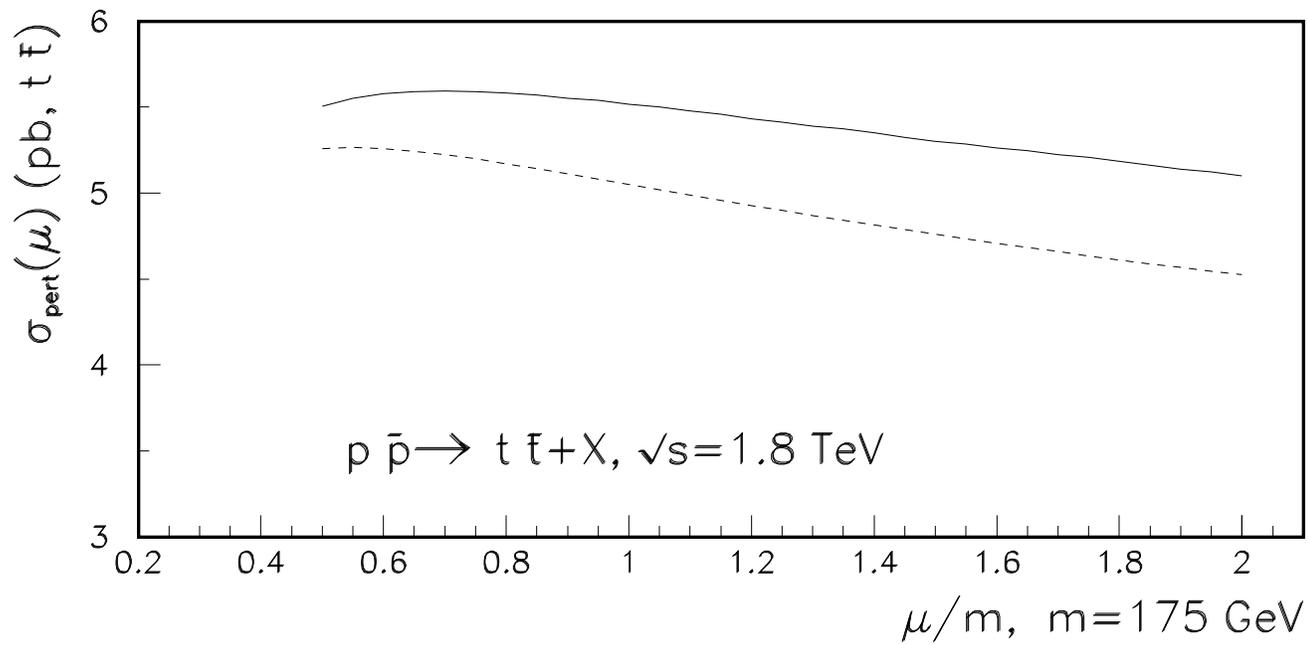
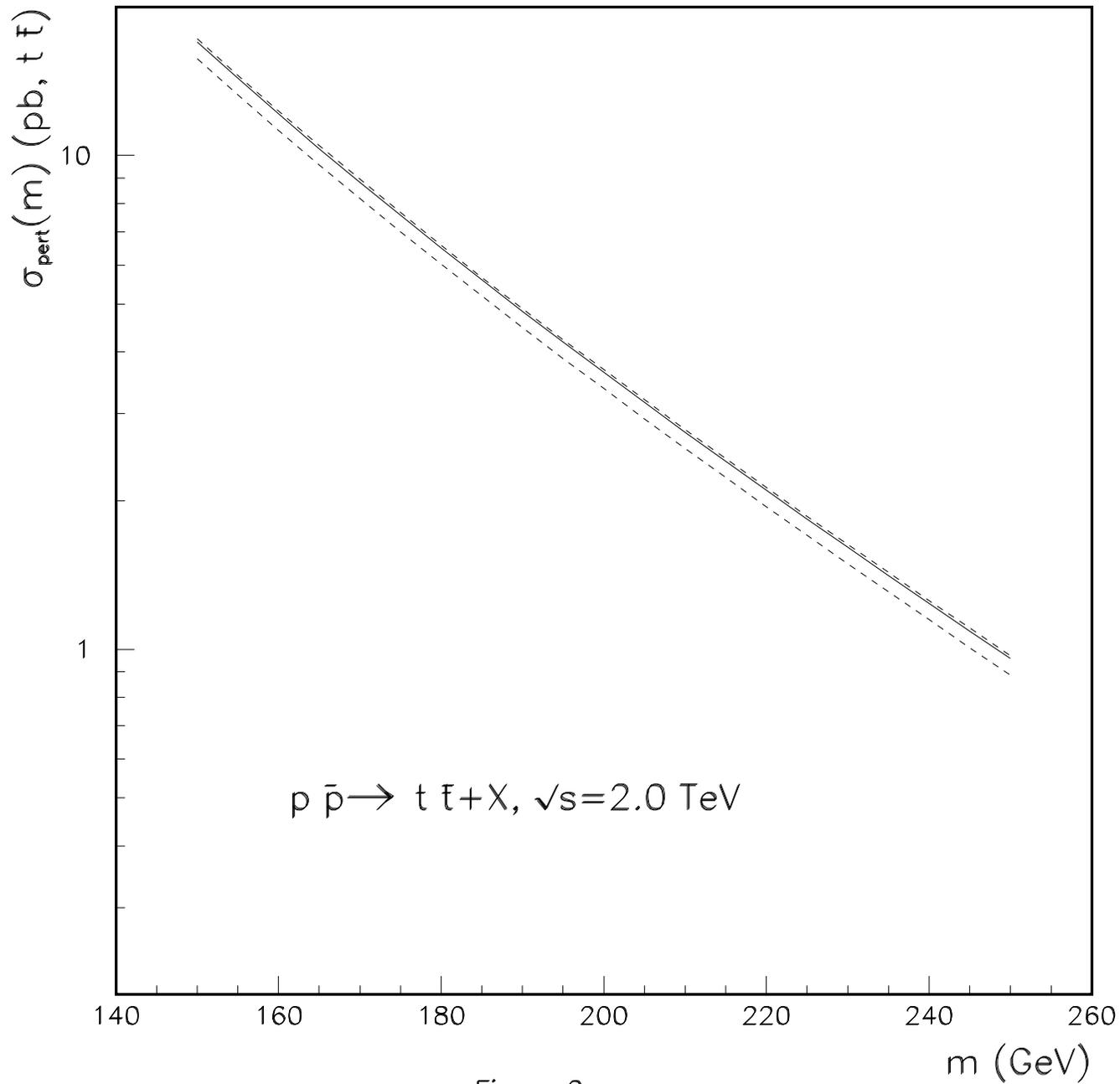
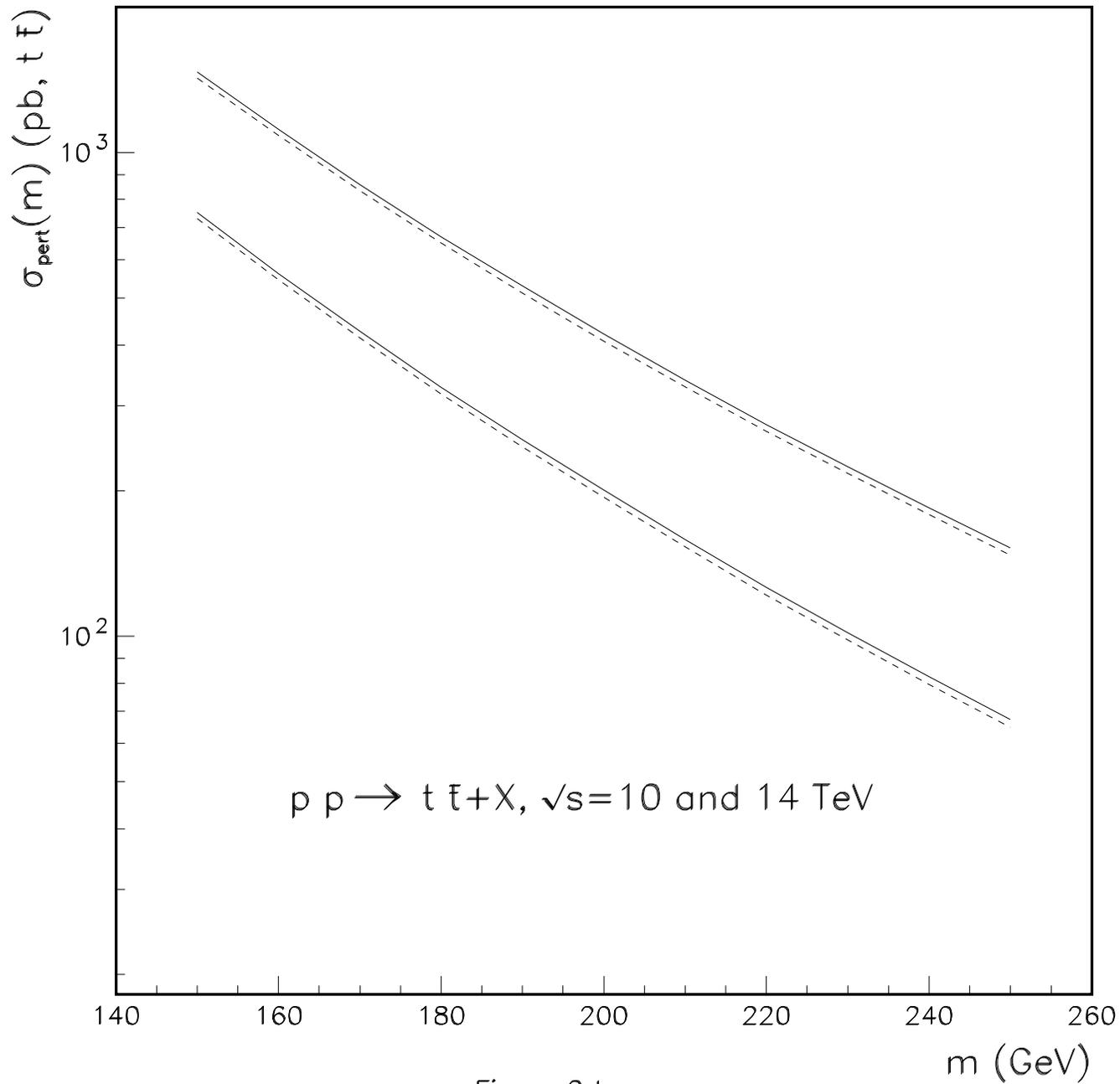
\end{document}